\documentclass[aps,prfluids,onecolumn,superscriptaddress,amsmath,amssymb]{revtex4-2}
\usepackage[utf8]{inputenc}   
\usepackage[T1]{fontenc}
\usepackage{lmodern}
\usepackage{bigints}
\usepackage[english]{babel}
\usepackage{graphicx}
\usepackage{bm}
\usepackage{comment}
\usepackage{overpic}
\usepackage{geometry}
\usepackage{color}
\usepackage{enumitem}
\usepackage{amsmath,amsfonts,amssymb}
\usepackage{subfigure}
\usepackage[subfigure]{tocloft}
\usepackage{hyperref} 

\usepackage[]{xcolor}

\newtheorem{prop}{Proposition}

\begin{document}

\title{Bending-compression coupling in extensible slender microswimmers}
\author{Kenta Ishimoto}\thanks{kenta.ishimoto@math.kyoto-u.ac.jp}
\affiliation{Department of Mathematics, Kyoto University, Kyoto 606-8502, Japan}
\author{Johann Herault}\thanks{johann.herault@imt-atlantique.fr}
\affiliation{Nantes Université, École Centrale Nantes, IMT Atlantique, CNRS, LS2N, UMR 6004, F-44000 Nantes, France}
\author{Cl\'ement Moreau}
\thanks{clement.moreau@ls2n.fr}
\affiliation{Nantes Université, École Centrale Nantes, IMT Atlantique, CNRS, LS2N, UMR 6004, F-44000 Nantes, France}
\date{\today}

\begin{abstract}
Undulatory slender objects have been a central theme in the hydrodynamics of swimming at low Reynolds number, where the slender body is usually assumed to be inextensible, although some microorganisms and artificial microrobots largely deform with compression and extension. Here, we theoretically study the coupling between the bending and compression/extension shape modes, using a geometrical formulation of microswimmer hydrodynamics to deal with the non-commutative effects between translation and rotation. 
By means of a coarse-grained minimal model and systematic perturbation expansions for small bending and compression/extension, we analytically derive the swimming velocities and report three main findings. 
First, we revisit the role of anisotropy in the drag ratio of the resistive force theory and generally demonstrate that no motion is possible for uniform compression with isotropic drag. 
We then find that the bending-compression/extension coupling generates lateral and rotational motion, which enhances the swimmer's manoeuvrability,
as well as changes in progressive velocity at a higher order of expansion, while the coupling effects depend on the phase difference between the two modes.
Finally, we demonstrate the importance of often-overlooked Lie bracket contributions in computing net locomotion from a deformation gait. Our study sheds light on compression as a forgotten degree of freedom in swimmer locomotion, with important implications for microswimmer hydrodynamics, including understanding of biological locomotion mechanisms and design of microrobots. 
\end{abstract}

\maketitle

\section{Introduction}

The hydrodynamics of swimming of slender bodies in the Stokes regime has been intensively studied in recent decades,
in particular in the context of swimming microorganisms with cilia and flagella and artificial microrobots made of rod-shaped particles and flexible filaments \cite{elgeti2015physics, lauga2020fluid}.
The locomotion of such swimmers is often achieved by undulatory deformation, and therefore studies on a bending filament in a viscous fluid have constituted a central theme in the field for more than a half century \cite{taylor1951analysis, gray1955propulsion}.
These slender objects are usually assumed to be inextensible, although this assumption may not be reasonable for some biological and artificial swimmers.

In fact, large compressions and extensions have been reported in some unicellular microorganisms.
A slender organelle, spasponeme, seen in {\it Vorticella} can rapidly contract within a second in response to calcium signalling \cite{ryu2016vorticella}, with its contraction being rather uniform.
A microtubule-based motile organelle, called haptonema, seen in Haptophyte algae, exhibits a rapid coiling contraction within milliseconds, while the detailed mechanisms are still unknown 
\cite{nomura2019microtubule}.
Microtubule-supported pseudopodia of Heliozoan, called axopodia, also exhibit contraction and extension \cite{suzaki1980food, febvre1986motility}.
Body contraction is well known in {\it Euglena} as euglenoid motion \cite{arroyo2012reverse} and such a body contraction is also seen in some ciliates such as  {\it Stentor} \cite{huang1973contractile}.  More recently, an origami-based mechanism has been uncovered for a large extensibility of a `neck' of {\it Lacrymaria} \cite{flaum2024curved}. 

The millimetre-sized nematode {\it Caenorhabditis elegans} is a well-studied model organism in a large domain of biology, including reproduction, development, and neuromechanics \cite{meneely2019working}, and its swimming dynamics are also well captured by fluid mechanics in the Stokesian regime \cite{montenegro2016flow}. Since the bending motion of {\it C. elegans} is driven by muscular contraction, the worm inevitably changes its body length during its movement. Hence, the 
measurement of compressibility has been performed experimentally, for example, by a linear viscoelastic model \cite{backholm2013viscoelastic}. The shape tracking through videomicroscopy also requires the body contraction/extension for accurate detection of the body shape \cite{roussel2014robust}. Integrated neuromechanical models of {\it C. elegans} therefore include body extensibility \cite{izquierdo2018head}.

In artificial microrobots, large compressions and extensions are well recognised in microswimmers made of hydrogels \cite{nikolov2015self, sharan2021fundamental, tan2024shape}. Motile extensible filament has been experimentally developed via a self-assembled droplet, in which self-elongation is essential to trigger morphological instability \cite{cholakova2021rechargeable, lisicki2024twist}.

The simplest model of the body-environment coupling of slender bodies in the inertialess regime is the resistive force theory (RFT), which only considers the local hydrodynamic drag and does not account for non-local hydrodynamic interactions \cite{gray1955propulsion, lighthill1976flagellar}. Nonetheless, because of its theoretical and computational simplicity, RFT has been widely used as an empirical drag model of viscous Newtonian fluid as well as non-Newtonian fluid \cite{fu2007theory, riley2017empirical} and  granular material \cite{zhang2014effectiveness}. More recently, RFT-type local drag models have been used even for terrestrial locomotion including slithering and legged animals and robots  \cite{zhang2014effectiveness, chong2023self, rieser2024geometric}. In these studies with RFT, net locomotion is generated by the drag anisotropy between the tangential and normal drag coefficients.

Despite these broad interests in the compressible/extensible slender microswimming from biological to artificial systems, our understanding about the impact of compression/extension when coupled to bending deformation is still very limited. 
Notable past theoretical studies on the role of compression/extension in microswimming deal with 
an extensible filament with isotropic drag ratio that moves in one dimension \cite{pak2011extensibility} and a slender helical body, numerically simulated with implicit extensibility by reparametrisation of the body curve \cite{pak2012hydrodynamics}. 
Another model featuring compression was studied in the context of crawling worm motions \cite{desimone2012crawling, tanaka2012mechanics}, which however focused on body-surface interactions with one-dimensional locomotion and does not contain a bending mode. Body extensibility was considered in simulations of an active filament in a viscous fluid but mainly to stabilise the numerical schemes \cite{olson2013modeling, ishimoto2018elastohydrodynamical}.

An exception is the two-link scallop swimmer with compression/extension, which has been recently studied by \cite{gidoni2024gait}, focusing on the controllability problem. 
The famous Purcell scallop theorem states \cite{purcell1977life, ishimoto2012coordinate} that a reciprocal deformation cycle cannot generate net locomotion in Stokes flow. 
Since compression/extension provides an additional degree of freedom, even the simple scallop model with a single bending angle can generate net locomotion. 
To the best of the authors' knowledge, however, the impacts of the compression/extension during bending motions have not been comprehensively studied.

Our aim in this study is therefore to theoretically elucidate the mechanical coupling between body compression/extension and bending of self-propelled slender bodies at low Reynolds number, which we hereafter refer to as the {\it bending-compression} coupling for brevity. For this purpose, we use the geometrical theory of microswimming \cite{shapere1989geometry}, which relies on the connection operator linking net motion as the product of a cyclic body deformation. The swimmer trajectories are obtained by a geometrical integration based on the Magnus expansion \cite{hatton2015nonconservativity},  with Lie brackets capturing the effects of motion non-commutativity.

Equipped with this framework, we make further analytical progress by two methods: coarse-graining the deformation of the swimmer to a minimal set of two degrees of freedom, and focusing on small-amplitude deformation in the general case. Regarding the choice of deformation cycles, we examine two illustrative examples, uniform compression and compression-bending wave, motivated by biological locomotion such as body contraction of unicellular microorganisms and muscular contraction of nematodes. 

Upon laying out the dynamics and deriving analytical expressions for the coarse-grained and small-amplitude cases, our results on the compression-bending dynamics in slender microswimmers may be outlined as three main findings. First, we demonstrate that compression-bending coupling allows net locomotion in the case of isotropic drag, even if the swimmer's total body length remains constant over time, rectifying a prior statement by \cite{pak2011extensibility} which overlooked the active role of local compression. Secondly, we observe the prominent role played by compression-bending coupling in enhancing the swimmer's ability to rotate in the fixed frame, through the emergence of a coupled lower-order term in the averaged rotational velocity from small amplitude theory. This improved turning ability is also confirmed by numerical simulations at large amplitude. Finally, the expressions we obtain for swimming velocity through perturbative analysis highlight the importance of the Lie bracket terms due to non-commutativity between rotation and translation in deriving the swimmer's net locomotion, which is often bypassed in analogous studies on flagellar swimming. The compression-bending swimmer provides an instructive example warranting the crucial role of non-commutative terms in a broader context of microswimming theory. 


The paper is organised as follows. In \S\ref{sec:problem_settings}, we introduce our general compressible/extensible slender object and present a general statement on locomotion with isotropic drag. In \S\ref{sec:minimal_model}, we consider a minimal theoretical model similar to the compressible/extensible scallop swimmer by \cite{gidoni2024gait}, focusing on elementary locomotion gaits and the case of isotropic drag. We then analyse a general slender object via a systematic perturbation theory on small-amplitude bending and compression in \S\ref{sec:small_amp}. Two example swimmers, uniform compression and compression-bending wave, are then examined in \S\ref{sec:uni_comp} and \S\ref{sec:comp_bend_wave}, respectively, followed by a discussion on swimmer manoeuvrability by compression in \S\ref{sec:manoeuvrability},
before final conclusions are drawn in \S\ref{sec:conc}.

\section{Problem setting and general properties}
\label{sec:problem_settings}

\subsection{Coordinates, kinematics and dynamics}

\begin{figure}[!t]
\begin{center}
\begin{overpic}[width=14cm]{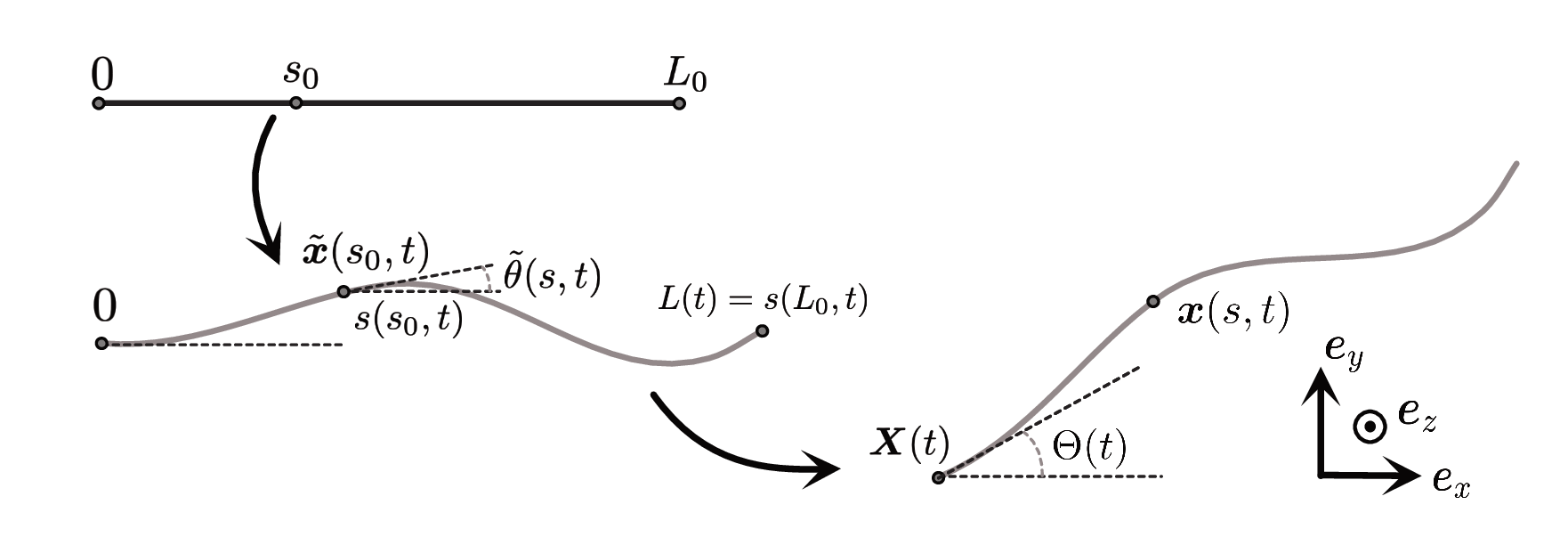}
\put(5,32){(a)}
\put(5,17){(b)}
\put(57,10){(c)}
\end{overpic}\\
\caption{Schematic of a filament in three frames of reference. (a) A filament in a reference state. The arc length $s_0\in [0, L_0]$ is used for parametrisation of the curve. (b) A filament in the body-fixed frame at time $t$. The point labelled by $s_0$ is mapped to $\tilde{\bm{x}}(s_0, t)$, where the distance along the filament is denoted by $s(s_0, t)$ and the local tangent angle is represented as $\tilde{\theta}(s_0, t)$. (c) A filament in the laboratory frame, which is obtained by rigid body transformation, translation with $\bm{X}(t)$ and rotation by $\Theta(t)$, from the filament in (b). The point $\tilde{\bm{x}}(s_0, t)$ is mapped to $\bm{x}(s,t)$.}
\label{fig:config}
\end{center}
\end{figure}

We consider a slender self-propelled object moving in a plane, which we denote as our $x-y$ plane, as illustrated in Figure~\ref{fig:config}. The length of the filament, $L$, may change over time due to self-compression/extension. To represent this effect, we introduce a Lagrangian label $s_0\in [0, L_0]$ as the arc length of the filament in the reference state, where $L_0$ is the reference length of the filament [Figure~\ref{fig:config}(a)]. We then consider the compression/extension at time $t$ in the form,
\begin{equation}
    \frac{\partial s}{\partial s_0}=1+\eta p(s_0, t), 
    \label{eq:def_p}
\end{equation}
where $s=s(s_0, t)$ is the arc length of the current configuration at time $t$ [Figure~\ref{fig:config}(b)]. Here, we introduce a constant $\eta$ for later use. In the body-fixed frame, the position labelled by $s_0$ is deformed to $\tilde{\bm{x}}(s_0, t)$ and the tangent angle is given by $\tilde{\theta}(s_0, t)$. We describe self-deformation of the swimmer as
\begin{equation}
    \tilde{\theta}(s_0, t)=\epsilon q(s_0, t)
    \label{eq:def_q},
\end{equation}
where the constant $\epsilon$ is introduced for later use.

The body shape in the laboratory frame is obtained by a rigid body transformation, translation with $\bm{X}(t)$ and rotation by $\Theta(t)$ as in Figure~\ref{fig:config}(c). The point $\tilde{\bm{x}}(s_0,t)$ in the body-fixed frame is then mapped to $\bm{x}(s, t)$ in the laboratory frame by $\bm{x}=\bm{X}+{\bf R}_\Theta\tilde{\bm{x}}$, where ${\bf R}_{\Theta}$ is the two-dimensional rotation matrix by an angle $\Theta$. In the kinematic problem considered in this study, we solve $\bm{X}(t)$ and $\Theta(t)$, provided the two intrinsic functions, $p(s_0, t)$ and $q(s_0, t)$.

The velocity of the local point on the body in the laboratory frame is obtained as the Lagrangian time derivative defined as $D \bm{x}/D t=\partial\bm{x}(s_0,t)/\partial t$. To compute this, we introduce the linear velocity $\bm{U}=(U_x, U_y)^\textrm{T}$ and the angular velocity $\bm{\Omega}=\Omega\bm{e}_z$ with $\Omega=\mathrm{d}\Theta/\mathrm{d}t$, 
where the superscript $\textrm{T}$ denotes the transpose. The local velocity is then given$\ $by 
\begin{equation}
    \frac{D \bm{x}}{D t}
    = \bm{U}+\Omega\bm{e}_z\times\tilde{\bm{x}}+{\bf R}_{\Theta}\frac{D{\tilde{\bm{x}}}}{D t}.
    \label{DXDt}
\end{equation}

To determine the motion of the object, we calculate the drag forces acting on an infinitesimal segment of the filament via the empirical resistive force theory. 
With $C_T(>0)$ and $C_N(>0)$ being the tangential and normal hydrodynamic drag coefficients per unit length, the drag from the environment is obtained as
\begin{equation}
    \bm{f}(s, t)=-C_{\|}\left(\frac{D\bm{x}}{D t} \cdot\bm{e}_{\|}\right)\bm{e}_{\|}-C_{\perp}\left(\frac{D\bm{x}}{D t} \cdot\bm{e}_{\perp}\right)\bm{e}_{\perp}
    \label{eq:RFT},
\end{equation}
where $\bm{e}_{\|}(s_0, t)$ and $\bm{e}_{\perp}(s_0, t)$ are tangential and normal unit vectors at the segment of the filament, respectively.
The ratio $\gamma:=C_N/C_T$ is called the anisotropy ratio and the particular case  $\gamma=1$ is called isotropic drag.
In this study, we take this as an arbitrary positive value. For swimming in low Reynolds number flow, it is known that $\gamma \rightarrow 2$ in the slender filament limit \cite{gray1955propulsion}, and empirically $\gamma=1.5-1.8$ is used for flagella and cilia \cite{lighthill1976flagellar, friedrich2010high}.
When the flagellum possesses an accessory hairy structure known as mastigonemes, however, the effective anisotropy ratio $\gamma$ may be lower than unity ($\gamma<1)$ \cite{brennen1976locomotion, asadzadeh2022hydrodynamic}. In contrast, locomotion of {\it C. elegans} on a gel-like structure is well described by a large anisotropy ratio of $\gamma\approx 70$ \cite{keaveny2017predicting}. 

To close the system and calculate the locomotion velocity, we employ the force-free, $\tilde{\bm{F}}=(\tilde{F}_x, \tilde{F}_y)^\textrm{T}=\bm{0}$, and torque-free conditions, $\tilde{\bm{M}}=M\bm{e}_z=\bm{0}$, in the body-fixed frame,
\begin{equation}
\int_0^{L(t)}\tilde{\bm{f}}(s, t)\,\mathrm{d}s=\int_0^{L(t)}\tilde{\bm{x}}\times\tilde{\bm{f}}(s, t)\,\mathrm{d}s=\bm{0},
\label{eq:balance}
\end{equation}
where we employ the torque around the point $\bm{X}(t)$.

We denote the translational velocity in the body-fixed frame by $\tilde{\bm{U}}=(\tilde{U}_x, \tilde{U}_y)^\textrm{T}$ and introduce a $3\times 3$ matrix representation of the Lie group $\mathcal{G}\in\textrm{SE}(2)$, also known as homogeneous representation, as
\begin{equation}
    \mathcal{G}=\begin{pmatrix}
        {\bf R}_{\Theta} & \bm{X} \\
        (0, 0)  & 1
    \end{pmatrix},
\end{equation}
and the associated Lie algebra $\mathrm{se}(2)$, whose elements $\hat{\mathcal{A}}$ act on $\mathrm{SE}(2)$ as  $\dot{\mathcal{G}}=\mathcal{G}\hat{\mathcal{A}}$, with the matrix representation
\begin{equation}
    \hat{\mathcal{A}}=\begin{pmatrix}
    0 & \Omega & \tilde{U}_x \\
   -\Omega & 0 &\tilde{U}_y \\    
   0 & 0 & 0
    \end{pmatrix}.
    \label{eq: hat A}
\end{equation} 
The ``hat'' notation of $\hat{\mathcal{A}}$ is standard in solid mechanics and refers to the correspondence between the matrix representation of the Lie algebra $\mathrm{se}(2)$ and its twist coordinates stacked in a vector (representation) $\mathcal{A} = (\hat{\mathcal{A}})^{\vee} = ( \tilde{U}_x,  \tilde{U}_y, \Omega )^\textrm{T} $. 

To derive the expression in the laboratory frame, we directly solve the differential equation on the rigid body motion 
and represent its solution as
\begin{equation}
    \mathcal{G}(T)=\mathcal{G}(0)\,\overline{\rm P}\exp\left[\int_0^T \hat{\mathcal{A}}(t)\,\mathrm{d}t \right],
    \label{eq:pathint}
\end{equation}
where the symbol $\overline{\textrm{P}}$ denotes the reverse path ordering operator \cite{shapere1989geometry}.  Using the Magnus expansion,  this formal solution, Eq.~\eqref{eq:pathint}, may be expanded in a Lie brackets series, as
\begin{multline}
    \mathcal{G}(T)=\mathcal{G}(0)\exp\bigg[\int_0^T \hat{\mathcal{A}}(t_1)\,\mathrm{d}t_1 
    +\frac{1}{2}\int_0^T\,\mathrm{d}t_1\int_0^{t_1}\,\mathrm{d}t_2 ~[\hat{\mathcal{A}}(t_1), \hat{\mathcal{A}}(t_2)] \\
    +\frac{1}{6}\int_0^T\!\!\!\mathrm{d}t_1\int_0^{t_1}\!\!\! \mathrm{d}t_2\int_0^{t_2}\!\!\! \mathrm{d}t_3 \big([\hat{\mathcal{A}}(t_1), [\hat{\mathcal{A}}(t_2), \hat{\mathcal{A}}(t_3)]+[\hat{\mathcal{A}}(t_3), [\hat{\mathcal{A}}(t_2), \hat{\mathcal{A}}(t_1)]\big)+\dots
    \bigg],
    \label{eq:magnus}
\end{multline}
where the bracket symbol represents the matrix commutator $[\hat{\mathcal{A}}(t_1), \hat{\mathcal{A}}(t_2)]=\hat{\mathcal{A}}(t_1)\hat{\mathcal{A}}(t_2)-\hat{\mathcal{A}}(t_2)\hat{\mathcal{A}}(t_1)$. 

An alternative path to the computation of $\mathcal{G}$ resorts to gauge field theory \cite{shapere1989geometry, zhao2022walking}.
By linearity of the resistive force theory, the matrix $\hat{\mathcal{A}}$ may be decomposed as
\begin{equation}
    \hat{\mathcal{A}}=\sum_\alpha \hat{\mathcal{H}}_\alpha \dot{\sigma}_\alpha,
    \label{eq: stokes connection}
\end{equation}
where $\sigma_\alpha$ ($\alpha=1,2,\cdots,N)$ is the shape variable and $N$ is the number of degrees of freedom in the shape space, with the dot symbol denoting the time derivative. Here, 
$\hat{\mathcal{H}}_\alpha$ is called a gauge field or Stokes connection, 
as it depends only on the shape variables. A local connection maps a shape variation to a velocity for a given body shape, independent of swimmer position and orientation \cite{kelly1995geometric}. Using the matrix representation of $\mathrm{SE}(2)$, the connection $\hat{\mathcal{H}}_{\alpha}$ is represented by a second-rank tensor of dimensions $3\times 3 \times N$; we may also ``drop the hats'' in Eq.~\eqref{eq: stokes connection}, using the twist coordinates $\mathcal{A}$, which would, in turn, represent the Stokes connection $\mathcal{H}_{\alpha}$ as a $3 \times N$ matrix  $\mathcal{H}=(\bm{H}_x; \bm{H}_y; \bm{H}_\theta)$, where the semicolon denotes vertical concatenation.
The integral representation [Eq.~\eqref{eq:pathint}] is then rewritten as
\begin{equation}
    \mathcal{G}(T)=\mathcal{G}(0)\overline{\textrm{P}}\exp\left[ \sum_\alpha\int_C \hat{\mathcal{H}}_\alpha(\bm{\sigma})\,\mathrm{d}\sigma_\alpha\right]
    \label{eq:path_integC},
\end{equation}
where $C$ is a path in the shape space.
We are particularly interested in the net locomotion caused by a periodic shape gait, which is represented by a closed loop in the shape space. In the small-amplitude case, by neglecting terms above third-order in the expansions, the net locomotion is well described by a surface integral of the curvature in the shape space as \cite{hatton2015nonconservativity}
\begin{equation}
    \mathcal{G}(T) \approx \mathcal{G}(0)\exp\left[\sum_{\alpha, 
    ~\beta}
    \int_{S_C} \hat{\mathcal{F}}_{\alpha\beta}(\bm{\sigma})\,\mathrm{d}\sigma_\alpha\,\mathrm{d}\sigma_\beta\right]
    \label{eq:gauge_veloc},
\end{equation}
where $S_C$ is a region in the shape space enclosed by the close loop $C$ and the associated curvature form, also called  field strength or Stokes curvature, is defined as
\begin{equation}
    \hat{\mathcal{F}}_{\alpha\beta}=\frac{\partial \hat{\mathcal{H}}_\alpha}{\partial \sigma_\beta}-\frac{\partial \hat{\mathcal{H}}_\beta}{\partial \sigma_\alpha}+[\hat{\mathcal{H}}_\alpha, \hat{\mathcal{H}}_\beta].
    \label{eq: curvature field}
\end{equation}
When pulled back to the vector representation of $\mathrm{se}(2)$, the curvature field $\mathcal{F}$ has $3$ components, $\mathcal{F}=(\mathcal{F}_x, \mathcal{F}_y, \mathcal{F}_{\theta})^{\textrm{T}}$, which can be interpreted as the effect of small loop gaits on each spatial direction according to Stokes' theorem. Finally, the averaged velocity is obtained as
\begin{equation}
    \langle\hat{\mathcal{A}}\rangle=\frac{1}{2}\sum_{\alpha, ~\beta}\hat{\mathcal{F}}_{\alpha\beta}\langle\sigma_\alpha \dot{\sigma}_\beta\rangle
    \label{eq:gauge_veloc},
\end{equation}
with the time average symbol, $\langle \bullet \rangle=(1/T)\int_0^T \bullet \,\mathrm{d}t$ and deformation time period $T$.

\subsection{Locomotion with isotropic drag}
\label{sec:locomotion-with-isotropic-drag}

In this section, we focus on the particular case of isotropic drag, i.e. $\gamma =1$. Note that the arguments below are not limited to planar motion, but are valid for general three-dimensional motion without external forces and torques.

Isotropic drag has been known to hinder locomotion. The classical RFT study by \cite{gray1955propulsion} considered simple sinusoidal transverse deformation and theoretically showed that drag anisotropy allows net locomotion.  
\cite{becker2003self} later extended this argument to a general self-propelled free-swimming slender object and showed that no net motion is possible in isotropic drag with only bending deformation. These studies, however, neglected rotational dynamics; it was then proven that isotropic drag may be compatible with net rotation \cite{koens2016rotation}. 
Finally, considering extensible slender swimmers, \cite{pak2011extensibility} argued that time variation of the total length can generate net locomotion in isotropic drag.  

However, ambiguity remains regarding the need for total length variation, and the exact role played by compression in allowing locomotion in isotropic drag. Here, we precisely answer with the proposition below. 

\begin{prop}
Assume that $\gamma=1$. Let $\overline{\bm{X}}$ be the centre of geometry of the swimmer, defined as
\begin{equation}
        \overline{\bm{X}}(t)=\frac{1}{L}\int_0^{L(t)} \bm{x}(s, t)\,\mathrm{d}s.
        \label{eq: center of mass}
    \end{equation}
Then, the following statements hold:
\begin{enumerate}[label=(\roman*)]
    \item \label{item : prop unif} If the compression is spatially uniform ($\partial p / \partial s_0 = 0$ for all times), then $\dot{\overline{\bm{X}}}=0$, i.e. no net motion is possible. 
    \item \label{item : prop non unif} Else ($\partial p / \partial s_0 \neq 0$), net motion is indeed possible. 
\end{enumerate}
\label{prop}
\end{prop}

Of particular note, case \ref{item : prop non unif} includes the case $\dot{L}=0$ of constant total length, contrary to the statement by \cite{pak2011extensibility}, which restrictively assumed that the arclength $s$ is a materially conserved quantity. This, however, is not always the case for active compression, as seen in the proof of Proposition \ref{prop} below.

\vspace{1em}

\textit{Proof of Proposition \ref{prop}}. Let us compute the time derivative of $\overline{\bm{X}}$, recalling that the arc length $s$ is provided as a function of the Lagrangian label $s_0$ as $s=s(s_0, t)$, which yields the following change of variable in the integral \eqref{eq: center of mass}:
    \begin{equation}
        \frac{\mathrm{d} \overline{\bm{X}}}{\mathrm{d}t}=
        \frac{\mathrm{d}}{\mathrm{d}t}\left[\frac{1}{L}\int_0^{L_0} \bm{x}(s_0, t)\left(1+\eta p(s_0, t)\right)\,\mathrm{d}s_0\right],
        \label{eq:time-evolv-COM}
    \end{equation}
    where $L_0$ is the body length in the reference frame.
    The right hand side of Eq.~\eqref{eq:time-evolv-COM} is then computed as
    \begin{eqnarray}
        -\frac{\dot{L}}{L^2}\int_0^L\bm{x}(s,t)\,\mathrm{d}s+\frac{1}{L}\int_0^{L_0}\frac{\partial \bm{x}(s_0, t)}{\partial t}\left(1+\eta p\right)\,\mathrm{d}s_0+\frac{\eta}{L} \int_0^{L_0}\bm{x}(s_0, t)\frac{\partial p}{\partial t}\,\mathrm{d}s_0.
        \label{eq: time derivative com}
    \end{eqnarray}
Assuming isotropic drag $(C_\|=C_\perp=C)$, the second term vanishes, hence one has
    \begin{equation}
        \int_0^{L_0}\frac{\partial \bm{x}(s_0, t)}{\partial t}(1+\eta p(s_0, t))\,\mathrm{d}s_0
        =\int_0^{L}\frac{D \bm{x}}{D t}\,\mathrm{d}s
        =\frac{1}{C}\int_0^{L}\bm{f}\,\mathrm{d}s=\bm{0}
    \end{equation}
    from the force balance equation.
    By writing $\dot{L}$ as 
    \begin{equation}
        \frac{\mathrm{d}L}{\mathrm{d}t}=\frac{\mathrm{d}}{\mathrm{d}t}\int_0^{L_0}\left[1+\eta p(s_0,t)\right]\,\mathrm{d}s_0=\eta\int_0^{L_0}\frac{\partial p(s_0, t)}{\partial t}\,\mathrm{d}s_0,
    \end{equation}
    we may summarise the motion of the centre of geometry in the isotropic drag case as
    \begin{eqnarray}
        \frac{\mathrm{d} \overline{\bm{X}}}{\mathrm{d}t}
        =\frac{\eta}{L}\left[ \int_0^{L_0}\left(\bm{x}(s_0, t)\frac{\partial p(s_0, t)}{\partial t}\right)\,\mathrm{d}s_0-
        \overline{\bm{X}}\int_0^{L_0}\frac{\partial p(s_0, t)}{\partial t}\,\mathrm{d}s_0\right]
        \label{eq:COM-iso-drag},
    \end{eqnarray}
    which is clearly nonzero in general, hence point \ref{item : prop non unif} of the Proposition. Note that we can also recover, from Equation \eqref{eq:COM-iso-drag}, the zero motion for an inextensible object, by setting $\eta=0$.

Now, we assume uniform compression, i.e. $\partial p/\partial s_0=0$, so $\partial p(s_0, t)/\partial t=\dot{p}$. Eq.~\eqref{eq:COM-iso-drag} then becomes
    \begin{equation}
\frac{\mathrm{d}\overline{\bm{X}}}{\mathrm{d}t}=\frac{\eta \dot{p}}{L}\left[ \int_0^{L_0}\bm{x}(s_0, t)\,\mathrm{d}s_0-L_0\overline{\bm{X}}\right]
        =\frac{\eta \dot{p}}{L}\left[ \frac{L_0}{L}\int_0^{L_0}\bm{x}(s_0, t)(1+\eta p)\,\mathrm{d}s_0-L_0\overline{\bm{X}}\right]
        =\bm{0},
    \end{equation}
which proves point \ref{item : prop unif}. $\square$

\vspace{1em}

The first term in the right-hand side of Equation \eqref{eq:COM-iso-drag} clearly highlights how the coupling between the local position and the compression rate contributes to the net motion, and this means in particular that a change of total length is not essential for the generation of motion. 
In the following sections, we quantitatively investigate the role of this coupling.

\section{Minimal model of swimming with bending-compression coupling}
\label{sec:minimal_model}

\begin{figure}
    \centering
    \includegraphics[width=0.65\linewidth]{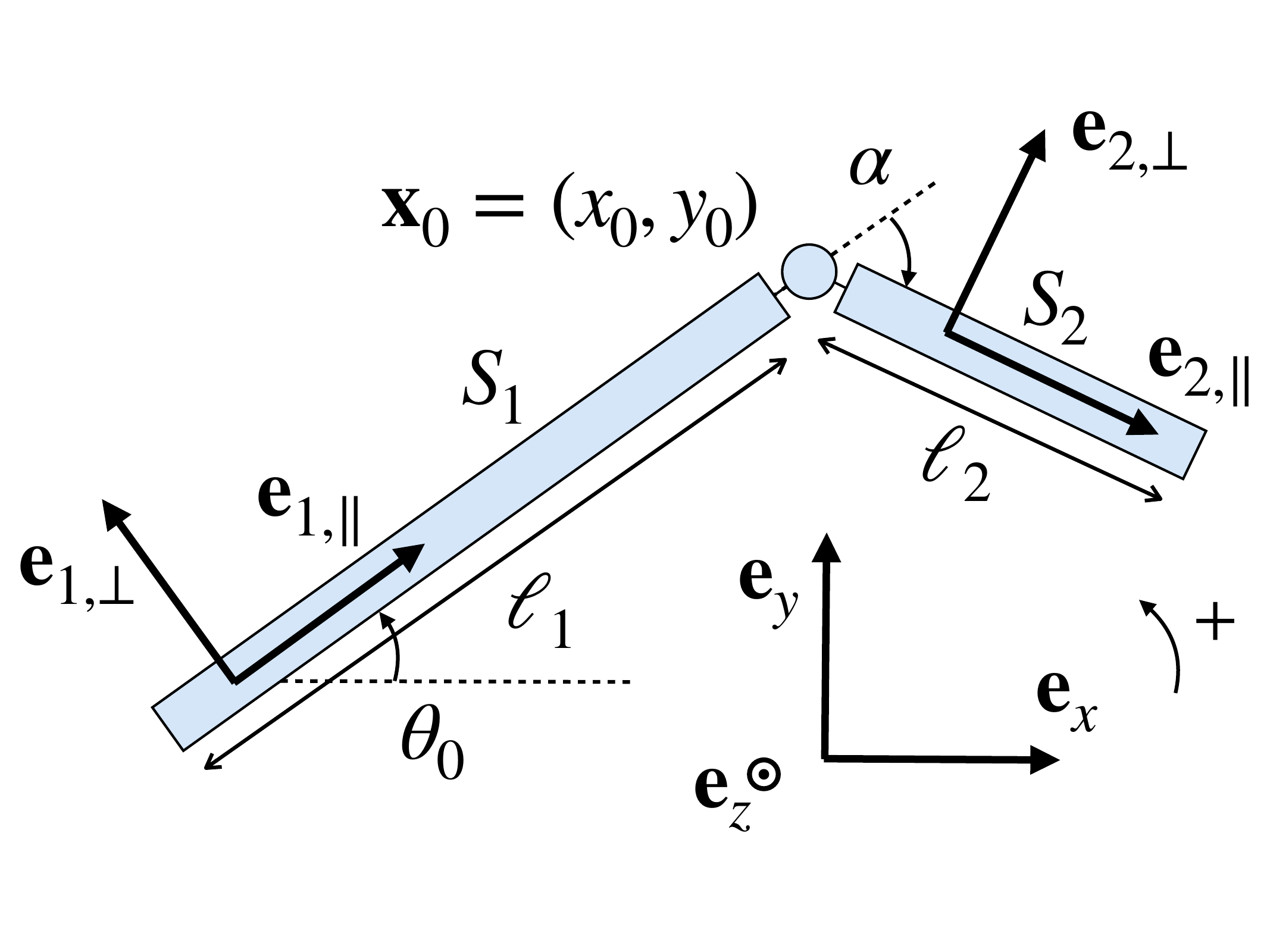}
    \vspace{-1.5em}
    \caption{Setup and notations for the minimal model.}
    \label{fig: setup minimal}
\end{figure}

In order to illustrate the statement of Proposition \ref{prop} and explore further how locomotion can result from the coupling between curvature-type and compression-type degrees of freedom,
we first consider a minimal model, made of two rigid links connected by a hinge (Figure~\ref{fig: setup minimal}). 
In this minimal model, compression is materialised by a uniformly distributed variation of the lengths of the links, while bending is represented by the varying angle between the links. 
Here, we consider an additional relationship between the lengths of the two links in order to retain only one independent parameter for compression; then, we obtain a system with two shape variables ($N=2$ in the previous section): one angle for bending, and one length ratio for compression.

If we further forbid compression, the model becomes the so-called ``scallop'', which famously cannot swim \cite{purcell1977life}: it is often used as an illustration of the fact that, in inertialess environments, at least two degrees of freedom are necessary for a deformation cycle to result in net locomotion, which can be seen from the connection structure in Eq.~\eqref{eq: stokes connection}.
In turn, several studies have proposed minimal models with a second degree of freedom, whether it is angular (``Purcell swimmer'') \cite{purcell1977life, moreau2019local, ishimoto2022self}, linear (``three-sphere swimmer'') \cite{najafi2004simple, yasuda2023generalized} or volumic (``push-me-pull-you swimmer'') \cite{avron2005pushmepullyou, silverberg2020realization}.

In this section, we add to this family what we could playfully call a ``squeeze-me-bend-you swimmer'', with compression as a second degree of freedom. A similar model has been studied in \cite{gidoni2024gait}, with a comparison between different distributions of compression along each link (i.e. the function $p(s_0,t)$), and focusing on model controllability with a limited range of $\gamma$. 
Beyond controllability, the simplicity of the model allows analytical computation of the Stokes connection and curvature fields, which we use here to gain intuition of the locomotion capabilities offered by compression-bending coupling, and propose elementary gaits.

\subsection{Model}

We consider an idealised planar swimmer made of two rigid slender rods $S_1$, $S_2$ of lengths $\ell_1$, $\ell_2$, connected by a junction and assume that the swimmer is situated in a $x-y$ plane. 
The position of the swimmer with respect to the fixed frame $(\bm{e}_x, \bm{e}_y, \bm{e}_z)$ is represented by the position of the junction between the two rods and is denoted by $\bm{x}_0 = (x_0,y_0)$. Additionally, we consider two moving frames $(\bm{e}_{1,\parallel},\bm{e}_{1,\bot})$ and $(\bm{e}_{2,\parallel},\bm{e}_{2,\bot})$ attached to each rod. We denote by $\theta_0$ the angle between $\bm{e}_{x}$ and $\bm{e}_{1,\parallel}$ and by $\alpha$ the angle between $\bm{e}_{1,\parallel}$ and $\bm{e}_{2,\parallel}$. These angles are taken positive for a counter-clockwise rotation from $\bm{e}_x$ and $\bm{e}_{1,\parallel}$, respectively.

Here, we assume that the angle $\alpha$ and the lengths $\ell_1$, $\ell_2$ evolve in time according to some prescribed gait, and aim to derive the effect of such a gait on the swimmer configuration in space: position $\bm{x}_0$ and orientation $\theta$. To track the current length of segment $S_i$ with respect to its reference length $\ell_i^0$, we introduce $\beta_i$ such that $\ell_i = \beta_i \ell_i^0$. 

In particular, we focus on two subcases of interest for the compression. The first case is {\it uniform compression}.
Unicellular swimming microorganisms such as {\it Stentor} and {\it Lacrymaria} exhibit morphological changes of their helically-coiled structure during the compression and extensions, and it is reasonable to assume uniform compression/extension as a simple model of their deformation.
In the two-link model, this means $\ell_1 =\ell_2$ at all times, or 
    \begin{equation}
        \beta_2 = \frac{\ell_1^0}{\ell_2^0} \beta_1
        \label{eq: rel beta1 beta2},
    \end{equation}
    which yields $\beta_1 = \beta_2$ at all times if $\ell_1^0 = \ell_2^0$. 
    
    The second subcase is the {\it constant total length}, which fits case \ref{item : prop non unif} of Proposition \ref{prop} and may model distributed muscular contraction. In the two-link model, it translates as $\ell_1 + \ell_2 = \ell_1^0 + \ell_2^0$ at all times, or
    \begin{equation}
        \beta_2 = \left ( 1 + \frac{\ell_1^0}{\ell_2^0} \right ) - \frac{\ell_1^0}{\ell_2^0} \beta_1
        \label{eq: rel beta1 beta2 2},
    \end{equation}
    which means $\beta_1 + \beta_2 = 2$ at all times, if $\ell_1^0 = \ell_2^0$.

For $s_0 \in [0,\ell_i^0]$, let $\bm{x}_{i}(s_0)$ be the coordinates of the point of Lagrangian label $s_0$ on segment $i$, with the origin of coordinates being taken at the $S_1-S_2$ junction. Then it holds that
\begin{align}
     \bm{x}_{1}(s_0) & = \bm{x}_0 - s_0 \beta_1 \bm{e}_{1,\parallel}, \\
     \bm{x}_{2}(s_0) & = \bm{x}_0 + s_0 \beta_2 \bm{e}_{2,\parallel},
\end{align}
and it follows, for the local velocity at $\bm{x}_{i}(s_0)$,
\begin{align}
    \dot{\bm{x}}_1 (s_0) & = \dot{\bm{x}}_0 - s_0 \dot{\beta}_1 \bm{e}_{1,\parallel} - s_0 \beta_1 \dot{\theta}  \bm{e}_{1,\bot} \\
    \dot{\bm{x}}_2 (s_0) & = \dot{\bm{x}}_0 + s_0 \dot{\beta}_2 \bm{e}_{2,\parallel} + s_0 \beta_2 (\dot{\theta} + \dot{\alpha})  \bm{e}_{2,\bot}.
\end{align}
Using equation \eqref{eq:RFT}, we can express the local force density exerted by the fluid on the swimmer at point $\bm{x}_{i}(s_0)$: 
\begin{equation}
\bm{f}_i (s_0) = - C_{\parallel} ( \dot{\bm{x}}_i \cdot \bm{e}_{\parallel} ) \bm{e}_{\parallel} - C_{\bot} ( \dot{\bm{x}}_i \cdot \bm{e}_{\bot} ) \bm{e}_{\bot}.
\end{equation}
Then, one can compute the total hydrodynamic force on each segment, scaled by extension $\beta_i$:
\begin{equation}
\bm{F}_i = \int_{S_i} \bm{f}_i \mathrm{d} S_i = \int_0^{\ell^0_i} \beta_i \bm{f}_i(s_0) \mathrm{d} s_0,
\end{equation}
which yields 
\begin{eqnarray}
    \bm{F}_1 &=& - \dot{x} \beta_1 \ell_1^0 (C_{\parallel} \cos \theta_0 \bm{e}_{1,\parallel} - C_{\bot} \sin \theta_0 \bm{e}_{1,\bot} )   \nonumber \\&&- \dot{y} \beta_1 \ell_1^0 (C_{\parallel} \sin \theta_0 \bm{e}_{1,\parallel} + C_{\bot} \cos \theta_0 \bm{e}_{1,\bot} )  \nonumber \\
    &&+ \dot{\theta}_0 C_{\bot} \beta_1^2 \frac{({\ell_1^0})^2}{2} \bm{e}_{1,\bot}
    + C_{\parallel} \beta_1 \dot{\beta}_1 \frac{({\ell_1^0})^2}{2} \bm{e}_{1,\parallel},
    \label{eq: force 1}\\
\bm{F}_2 &=& - \dot{x}_0 \beta_2 \ell_2^0 \, (C_{\parallel} \cos (\theta_0 + \alpha) \bm{e}_{2,\parallel} - C_{\bot} \sin (\theta_0 + \alpha) \bm{e}_{2,\bot} )  \nonumber \\ 
&& - \dot{y}_0 \, \beta_2 \ell_2^0 \, (C_{\parallel} \sin (\theta_0 + \alpha) \bm{e}_{2,\parallel} + C_{\bot} \cos (\theta_0 + \alpha) \bm{e}_{2,\bot} ) \nonumber \\ 
&&+ ( \dot{\theta}_0 + \dot{\alpha} ) C_{\bot} \beta_2^2 \frac{({\ell_2^0})^2}{2} \bm{e}_{2,\bot} + C_{\parallel} \beta_2 \dot{\beta}_2 \frac{({\ell_2^0})^2}{2} \bm{e}_{2,\parallel}.
\end{eqnarray}
Similarly, the torques $T_1$ and $T_2$ on each segment with respect to the junction point, and projected on $\bm{e}_z$, are given by
\begin{align}
    T_1 & = \beta_1^2 C_{\perp} \frac{({\ell_1^0})^2}{2}  \left ( - \dot{x}_0 \cos \theta_0 + \dot{y}_0 \cos \theta_0 \right ) + \dot{\theta}_0 \beta_1^3 \frac{({\ell_1^0})^3}{3}, \\
    T_2 & = \beta_2^2 C_{\perp} \frac{({\ell_2^0})^2}{2}  \left ( \dot{x}_0 \cos( \theta_0 + \alpha) + \dot{y}_0 \cos( \theta_0 + \alpha) \right ) + (\dot{\theta}_0 + \dot{\alpha}) \beta_2^3 \frac{({\ell_2^0})^3}{3}.
    \label{eq: torque 2}
 \end{align}
Since inertia is negligible, we can write balance of force and torques at all times:
\begin{equation}
    \left \{ 
    \begin{array}{l}
         \bm{F}_1 + \bm{F}_2 = 0, \\
         T_1 + T_2 = 0.
    \end{array}
    \right.
    \label{eq: force torque balance}
\end{equation}
 
Using equations \eqref{eq: force 1} to \eqref{eq: force torque balance}, and plugging the expression of $\beta_2$ with respect to $\beta_1$, one can derive the Stokes connection which links shape velocities $(\dot{\alpha},\dot{\beta}_1)$ to rigid motion velocities in the reference frame $(\dot{x}_0,\dot{y}_0,\dot{\theta}_0)$:
\begin{equation}
    \mathbf{A}(\theta_0,\alpha,\beta_1)\begin{pmatrix} \dot{x}_0 \\ \dot{y}_0 \\ \dot{\theta}_0 \end{pmatrix} = \mathbf{B} (\theta_0,\alpha,\beta_1) \begin{pmatrix} \dot{\alpha} \\ \dot{\beta_1} \end{pmatrix}.
    \label{eq: connection}
\end{equation}

To highlight the existence of a connection, we rewrite the resistance matrices in a moving frame $(\bm{e}_{a,\parallel},\bm{e}_{a,\bot})$ representing the average orientation of the swimmer, namely at $\theta_0 + \alpha/2$. To do so, we define 
$\mathbf{A}_a = \mathbf{R}_{\theta_0+\alpha/2}^{-1}  \mathbf{A} \mathbf{R}_{\theta_0+\alpha/2}$ and $\mathbf{B}_a = \mathbf{R}_{\theta_0+\alpha/2}^{-1} \mathbf{B}$, and write Eq.~\eqref{eq: connection} in the form,
\begin{equation}
    \mathbf{R} ^{-1}_{\theta_0+\alpha/2} \begin{pmatrix} \dot{x}_0 \\ \dot{y}_0 \\ \dot{\theta}_0 \end{pmatrix} = \mathcal{H}\begin{pmatrix} \dot{\alpha} \\ \dot{\beta_1} \end{pmatrix},
\end{equation}
where the connection $\mathcal{H}$ is given by $\mathcal{H}=\mathbf{A}_a^{-1} \mathbf{B}_a$.

Now, let us give detailed expressions of these matrices in both cases of interest \eqref{eq: rel beta1 beta2} and \eqref{eq: rel beta1 beta2 2} For simplicity, we assume $\ell^0_1 = \ell^0_2 = \ell$. In the case of uniform compression \eqref{eq: rel beta1 beta2}, one obtains
\begin{equation}
\mathcal{H}  =
\begin{pmatrix}
0 & 0\\
-\cfrac{\beta_1 \,\gamma \,\ell \,\cos \left(\frac{\alpha }{2}\right)}{4\,{\left[1+(\gamma-1) \,{\cos^2 \left(\frac{\alpha}{2}\right)}\right]}} & \cfrac{\ell \,\sin \left(\frac{\alpha }{2}\right)}{2\,{\left[\gamma+(1-\gamma) \,{\sin^2\left( \frac{\alpha }{2}\right)}\right]}}\\
-1/2 & 0
\end{pmatrix}.
\end{equation}
Then, one can compute the velocity of the geometric centre from its definition as 
$
    \bm{x}_m = \frac{1}{4} \left ( 2 \bm{x}_0 + \bm{x}_1(\ell) + \bm{x}_2(\ell) \right ),
$
which, expressed in the rotating basis $(\bm{e}_{a,\parallel},\bm{e}_{a,\bot})$, yields the motion only in the $y_m$ direction with its velocity given by
\begin{eqnarray}
    \dot{y}_m=
    \displaystyle \frac{\ell \,\cos \left(\frac{\alpha }{2}\right)\,{\left(\gamma -1\right)}\,{\left[\mathrm{\dot{\beta}_1}\,\sin \alpha 
+ \frac{\mathrm{\dot{\alpha}}\,\beta_1 }{2}(\cos \alpha -1)
\right]}}{2\,\gamma -2\,\cos \alpha +2\,\gamma \,\cos \alpha +2},
\end{eqnarray}
from which we deduce that any translational motion occurs along the $y_m$ direction. Moreover, it is clear that isotropic drag ($\gamma = 1$) forbids any translation, in agreement with Case \ref{item : prop unif} of Proposition \ref{prop}. 

In the case of constant total length \eqref{eq: rel beta1 beta2 2}, one has
\begin{equation}
\mathbf{A}_a = 
\left(\begin{array}{ccc}
\scriptstyle \ell \,{\left(\gamma +\cos \alpha -\gamma \,\cos \alpha +1\right)} &
\scriptstyle \ell \,\sin \alpha \,{\left(\beta_1 -1\right)}\,{\left(\gamma -1\right)} &
\scriptstyle -\gamma \,{\ell }^2 \,\sin \left(\alpha/2\right)\,{\left({\beta_1 }^2 -2\,\beta_1 +2\right)} \\
\scriptstyle \ell \,\sin \alpha \,{\left(\beta_1 -1\right)}\,{\left(\gamma -1\right)} &
\scriptstyle \ell \,{\left(\gamma -\cos \alpha+\gamma \,\cos \alpha+1\right)} &
\scriptstyle -2\,\gamma \,{\ell }^2 \,\cos \left(\alpha/2\right)\,{\left(\beta_1 -1\right)} \\
\scriptstyle -\gamma \,{\ell }^2 \,\sin \left(\alpha/2\right)\,{\left({\beta_1 }^2 -2\,\beta_1 +2\right)}  &
\scriptstyle -2\,\gamma \,{\ell }^2 \,\cos \left(\alpha/2\right)\,{\left(\beta_1 -1\right)}  &
\scriptstyle-(2/3)\,\gamma \,{\ell }^3 \,{\left(3\,{\beta_1 }^2 -6\,\beta_1 +4\right)}
\end{array}\right),\\
\mathrm{}\\
\end{equation}
and
\begin{equation}
\mathbf{B}_a =  \begin{pmatrix}
-(\gamma/2) \,{\ell }^2 \,\sin \left(\alpha/2\right)\,{{\left(\beta_1 -2\right)}}^2  & {\ell }^2 \,\cos \left(\alpha/2\right)\\
(\gamma/2) \,{\ell }^2 \,\cos \left(\alpha/2\right)\,{{\left(\beta_1 -2\right)}}^2 & -{\ell }^2 \,\sin \left(\alpha/2\right)\,{\left(\beta_1 -1\right)}\\
(\gamma/3) \,{\ell }^3 \,{{\left(\beta_1 -2\right)}}^3  & 0
\end{pmatrix},
\end{equation}
leading to the connection $\mathcal{H} = \mathbf{A}_a^{-1} \mathbf{B}_a$ in the form
\begin{equation}
\mathcal{H}   = \frac{1}{Q(\cos \alpha)} \begin{pmatrix} 2 \gamma \ell \beta_1^2 (\beta_1-2)^2 (\beta_1-1) \sin (\alpha/2) P_{11} (\alpha) &  \ell \cos (\alpha/2) P_{12} (\alpha) \\
\gamma \ell \beta_1^2 (\beta_1-2)^2 \cos (\alpha/2) P_{21} (\alpha) & \gamma \ell (\beta_1-1)  \sin (\alpha/2) P_{22}(\alpha) \\
(\beta_1-2)^2 P_{31}(\alpha) &  \beta_1 (\beta_1-2) P_{32}(\alpha) \end{pmatrix},
\end{equation}
where $P_{ij}$ are polynomials in $\cos \alpha$, $\sin \alpha$, $\beta_1$, whose expressions are too lengthy to write fully, and $\gamma$, and $Q$ is defined as
\begin{multline}
    Q(x) = {l_1 }^4 \,\big [ 3 \beta_1^4 (7 \,\gamma^2 \,{x}^2 -7 \,\gamma^2 -11 \,\gamma \,{x}^2 -2 \,\gamma \,x+13 \,\gamma +4 \,{x}^2 -4) \\
    + 12 \beta_1^3 (- 7 \,\gamma^2 \,{x}^2 +7 \,\gamma^2 +11 \,\gamma \,{x}^2 +2 \,\gamma \,x-13 \,\gamma -4 \,{x}^2 +4) \\
    + 16 {\beta_1 }^2 ( 7 \,\gamma^2 \,{x}^2 - 7 \, \,\gamma^2 - 11 \,\gamma \,{x}^2 - \frac{3}{2} \,\gamma \,x+ \frac{37}{2} \,\gamma +4 \,{x}^2 -4) \\
    + 8 \beta_1 ( - 7 \,\gamma^2 \,{x}^2 + 7 \,\gamma^2 +11 \,\gamma \,{x}^2 - 35 \,\gamma -4 \,{x}^2 +4) +112\,\gamma \big ].
\end{multline}
The expression for the motion of the geometric centre can be calculated with symbolic calculus software but is too tedious to be reproduced here, even in the particular case of isotropic drag. We can, however, evaluate it at specific shapes in the case $\gamma = 1$ to ensure motion is indeed possible. For isotropic compression ($\gamma = 1$) and flat shape ($\alpha = 0, \theta_0 = 0$), one obtains
\begin{equation}
    \dot{x}_m=\frac{\,\ell }{2} \dot{\beta_1}~,~~
    \dot{y}_m=-\frac{7\,\,\beta_1 \,\ell \,{{\left(\beta_1 -1\right)}}^2 \,{\left(\beta_1 -2\right)}}{48\,{\beta_1 }^2 -96\,\beta_1 +56} \dot{\alpha},
\end{equation}
and for links of equal lengths ($\beta_1 =1$), one obtains 
\begin{equation}
    \dot{x}_m=\frac{\,\ell \cos(\alpha + \theta_0)}{2} \dot{\beta_1}~,~~
    \dot{y}_m=\frac{\,\ell \sin(\alpha + \theta_0)}{2} \dot{\beta_1}.
\end{equation}
In both cases, it is clear that the geometric centre does not stay fixed, despite the total length of the swimmer remaining constant. This simple example therefore provides a constructive proof for point \ref{item : prop non unif} of Proposition \ref{prop}.

\subsection{Gait prediction}

From the expression of the connection, one may obtain the curvature of the gauge field $\mathcal{F}$ following the formula \eqref{eq: curvature field}. Since the squeeze-me-bend-you swimmer possesses only two degrees of freedom, each scalar component of $\mathcal{F}$ can be easily visualised on a two-dimensional plane $(\alpha,\beta)$. Moreover, any deformation gait can be represented in the same plane, and formula \eqref{eq: stokes connection} indicates that the net locomotion produced by the gait will be, at first order, proportional to $\mathcal{F}$ integrated over the (algebraic) area enclosed by the gait cycle in the shape space. In the following, we use this principle to design simple gaits achieving elementary motion in each available direction, demonstrating the swimming capabilities of the squeeze-me-bend-you swimmer model.

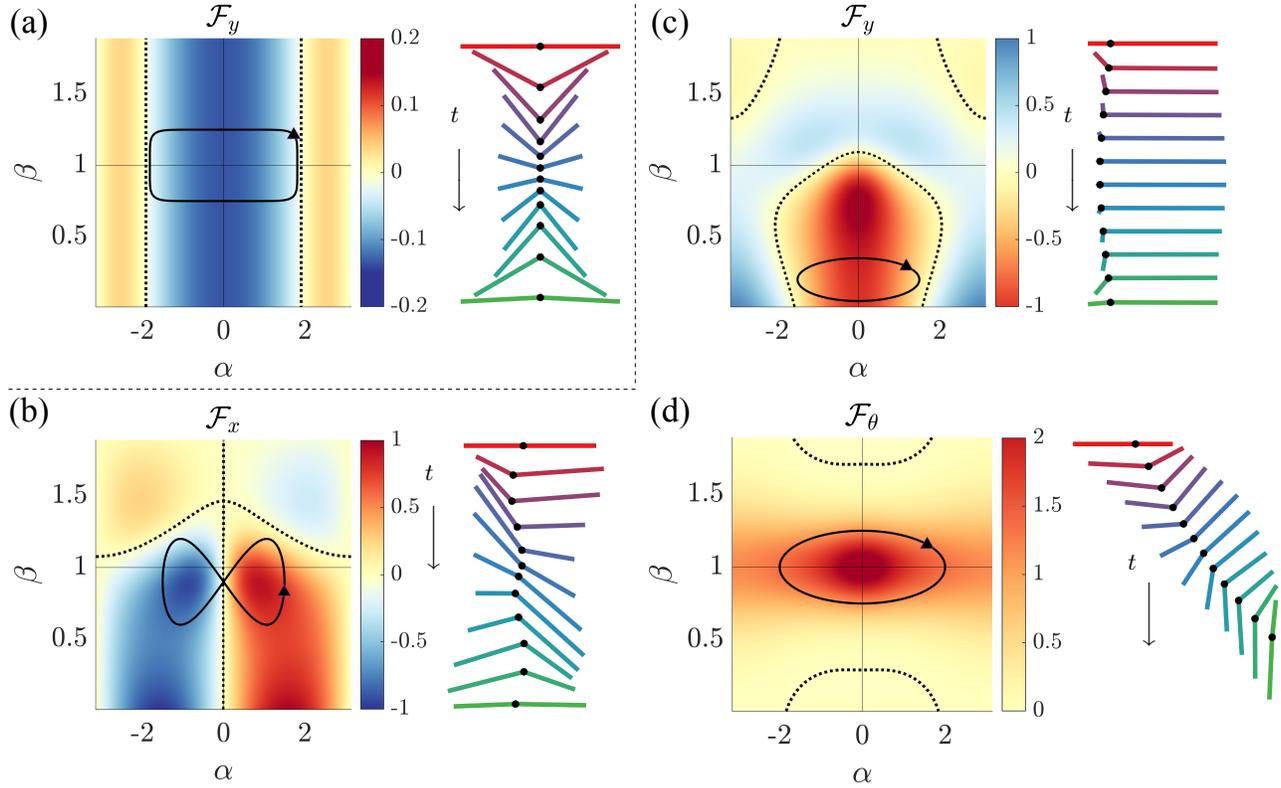
\begin{figure}
    \centering
    \begin{overpic}[width=\linewidth]{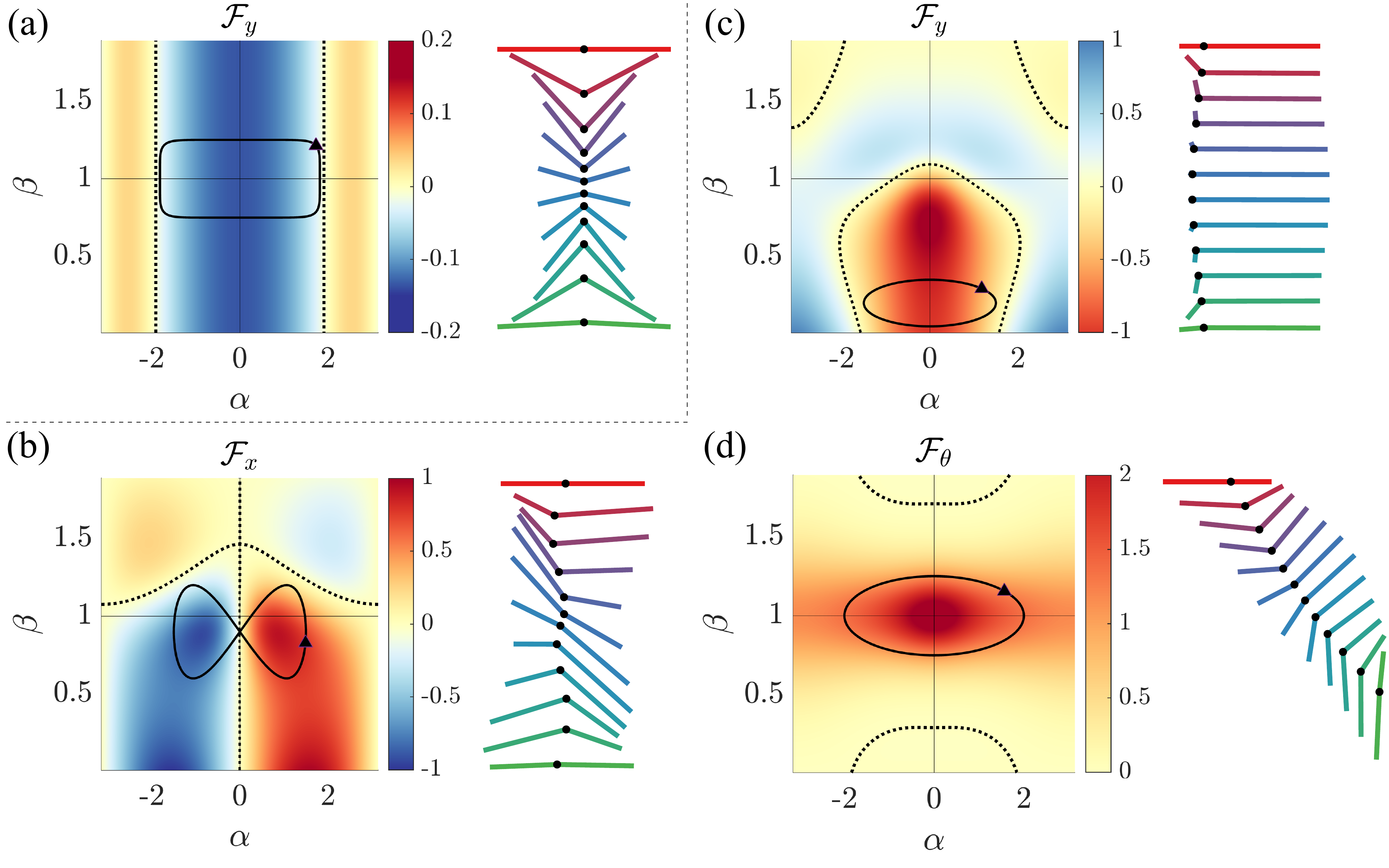}
    \put(35, 48){$\bigg\downarrow$}
    \put(33, 20){$\bigg\downarrow$}
    \put(83, 48){$\bigg\downarrow$}
    \put(89, 14){$\bigg\downarrow$}
    \put(35, 53){$t$}
    \put(33, 25){$t$}
    \put(83, 53){$t$}
    \put(88, 18){$t$}
    \end{overpic}    
    \caption{Curvature fields for the squeeze-me-bend-you swimmer, as defined in Equation \eqref{eq: curvature field}. Dotted black lines indicate zero-curvature level set. Suggested gaits are indicated by a continuous black line, and the corresponding deformation sequence of the swimmer is represented on the right of each curvature plot : (a) $y$-displacement for uniform compression, (b) $x$-displacement, (c) $y$-displacement, and (d) $\theta$-displacement with time progress shown by the arrow and colours.}
    \label{fig: curvature fields}
\end{figure}

We first examine the uniform compression case, in which only the $y$ component of $\mathcal{F}$ does not vanish, and takes a remarkably simple form depending only on $\alpha$ (and notably not on $\beta$):
\begin{equation}
    \mathcal{F}_y = \frac{\cos \left(\frac{\alpha }{2}\right)-3\,{\cos \left(\frac{\alpha }{2}\right)}^3 }{{{\left(2\,{\cos \left(\frac{\alpha }{2}\right)}^2 +2\right)}}^2 }.
\end{equation}
The function $\mathcal{F}_y$ is even and $2 \pi$-periodic. On the interval $[-\pi,\pi]$, it vanishes at $\pm \pi$ and $\pm \alpha_m$ with $\alpha_m= \arccos\left (- 1/3\right )$ (which, serendipitously, happens to be ``tetrahedral angle'' measured between a diagonal of a cube and its adjacent edge). It is therefore of constant sign on the interval $[-\alpha_m,\alpha_m]$, suggesting that, to ensure maximal displacement, the swimmer must fold between these two values, with compression $\beta$ alternating between any two set values with the same periodicity in the meantime. Such a gait is represented in Figure \ref{fig: curvature fields}(a).

In the constant length case, the symbolic expression of $\mathcal{F}$ is not nearly nice enough to be analysed in the same way, or even reproduced here. We may nonetheless evaluate it numerically, and its three components are plotted in Figure \ref{fig: curvature fields}(b)-(d). 
In Figure \ref{fig: curvature fields}(b), one may first observe that $\mathcal{F}_x$ is odd in $\alpha$. For this reason, a simple circular or elliptical gait centred at $(0,1)$ would fail to produce any net locomotion in the direction $x$. A good candidate for a successful gait would be a ``butterfly'' gait as shown in Figure \ref{fig: curvature fields}(b). Note that, in the meantime, no motion in $y$ or $\theta$ would result from this butterfly gait, due to the even parity in $\alpha$ of $\mathcal{F}_y$ and $\mathcal{F}_{\theta}$ visible in Figure \ref{fig: curvature fields}(c)-(d).
On the other hand, to generate net motion purely in the $y$ direction, an elliptical gait centred at $\alpha = 0$ may be considered, but needs to be shifted away from $\beta = 1$ to avoid producing net rotation at the same time, as shown in Figure \ref{fig: curvature fields}(b). 

As one can infer from $\mathcal{F}_{\theta}$ in Figure \ref{fig: curvature fields}(d), elliptical gaits centred at $(0,1)$ will rotate the swimmer in the reference plane. We note that the net rotation after one cycle is outstandingly high when compared to the capabilities of its cousin in the family of minimal swimmer models, the Purcell swimmer, which features only bending and no compression. Indeed, under the reasonable assumption that  $\alpha \in [-\pi,\pi]$, a simple optimisation algorithm on elliptical gaits shows that the Purcell swimmer can reach a net rotation of around $\pi/2$ rad in one cycle, at most, while the squeeze-me-bend-you swimmer can achieve a rotation of nearly $3\pi/2$. This high orientational manoeuvrability suggests that compression, even in a small amount, plays a prominent role in the orientational dynamics of slender swimmers. We further quantitatively evaluate this enhancement of manoeuverability in Section \ref{sec:manoeuvrability}.

\section{Small-amplitude theory}
\label{sec:small_amp}

In this section, we go back to a general slender object, but with a particular focus on a small bending and compression regime in order to quantify the effects of bending-compression coupling on the locomotion at low Reynolds number, by assuming  $\epsilon, \eta \ll 1$ and $\eta=O(\epsilon)$.

The position in the body-fixed frame is then expanded such that
\begin{eqnarray}
    \tilde{x}(s_0, t)&=&\int_0^{s_0}\cos(\theta(s_0', t))(1+\eta p(s_0',t))\,\mathrm{d}s_0'\nonumber\\
    &=& s_0+\eta\int_0^{s_0}p(s_0',t)\,\mathrm{d}s_0'-\frac{\epsilon^2}{2}\int_0^{s_0} q^2(s_0',t)\,\mathrm{d}s_0'+O(\eta\epsilon^2)
    \label{eq:tilde_x},\\
    \tilde{y}(s_0, t)&=&\int_0^{s_0}\sin(\theta(s_0', t))(1+\eta p(s_0',t))\,\mathrm{d}s_0' \nonumber \\
    &=&
    \epsilon\int_0^{s_0}q(s_0',t)\left[ 1+\eta p(s_0',t) \right]\,\mathrm{d}s_0'+O(\epsilon^3) \label{eq:tilde_y}.
\end{eqnarray}
Also, the local tangent and normal vectors need to be expanded as
\begin{eqnarray}
    \bm{e}_{\|}(s_0, t)=(1, \epsilon q)^\textrm{T}+O(\epsilon^2)~\textrm{and}~
    \bm{e}_{\perp}(s_0, t)=(-\epsilon q, 1)^\textrm{T}+O(\epsilon^2).
\end{eqnarray}
By plugging Eqs.~\eqref{eq:tilde_x}-\eqref{eq:tilde_y} into \eqref{DXDt}, we have the deformation velocity as
\begin{eqnarray}
    \frac{D\tilde{x}}{Dt}&=&\frac{\partial\tilde{x}}{\partial t}(s_0, t)
    =\eta\int_0^{s_0}\frac{\partial p}{\partial t}(s_0', t)\,\mathrm{d}s_0'-\frac{\epsilon^2}{2}\int_0^{s_0} \frac{\partial q^2(s_0',t)}{\partial t}\,\mathrm{d}s_0'+O(\eta\epsilon^2) \\
    \frac{D\tilde{y}}{Dt}&=&\frac{\partial\tilde{y}}{\partial t}(s_0, t)    =\epsilon\int_0^{s_0}\frac{\partial }{\partial t}\bigg[q(s_0',t)\left( 1+\eta p(s_0',t)\right) \bigg]\,\mathrm{d}s_0'+O(\epsilon^2), 
\end{eqnarray}
noting that the perpendicular deformation velocity contains both the bending and compression.

We consider expansions with respect to the small parameters $\epsilon$ and $\eta$ for the velocities in the body-fixed frame and rotational velocity. With the velocities in the body-fixed frame written as
$\tilde{\bm{U}}=(\tilde{U}_x, \tilde{U}_y)^\textrm{T}$, i.e., $\bm{U}={\bf R}_{\Theta}\tilde{\bm{U}}$, we use the following notations for the expansions:
\begin{eqnarray}
    \tilde{U}_x&=&\tilde{U}^{(1)}_x+\tilde{U}^{(2)}_x+\tilde{U}^{(3)}_x+\dots, \\
    \tilde{U}_y&=&\tilde{U}^{(1)}_y+\tilde{U}^{(2)}_y+\tilde{U}^{(3)}_y+\dots, \\
    \Omega&=&\Omega^{(1)}+\Omega^{(2)}+\Omega^{(3)}+\dots,
\end{eqnarray}
where the superscripts $^{(1)}$, $^{(2)}$, and $^{(3)}$ indicate the linear, quadratic, and cubic terms of $\epsilon$ and $\eta$, respectively.

\subsection{Leading-order calculations}

We first consider $\tilde{F}_x=0$ by substituting Eqs.~\eqref{eq:def_p}-\eqref{eq:def_q} into Eq.~\eqref{eq:RFT}, to obtain, up to the second order of expansions,
\begin{eqnarray}
    \int_0^L\left(\tilde{U}_x+\frac{D\tilde{x}}{Dt}\right)\,\mathrm{d}s=
    \epsilon\Omega\int_0^L q\tilde{y}\,\mathrm{d}s
    +\epsilon (\gamma-1) \int_0^L q\left(\tilde{U}_y+\Omega\tilde{x}+\frac{D\tilde{y}}{Dt}\right)ds+O(\epsilon^3, \epsilon^2\eta)
    \label{eq:Fx}.
\end{eqnarray}
Here, we find that the right-hand side of Eq.~\eqref{eq:Fx} is of the second order of small parameters, while the left-hand side is of the first order of magnitudes. Hence, at the leading order of the expansions, we have
\begin{eqnarray}
    \tilde{U}^{(1)}_x
    =-\frac{\eta}{L}\int_0^{L_0}\left[\int_0^{s_0}\frac{\partial p}{\partial t}(s'_0, t)\,\mathrm{d}s'_0\right] \mathrm{d}s_0.
    \label{eq:Ux1}
\end{eqnarray}

We then consider the force balance in the normal direction by imposing $\tilde{F}_y=0$, which is written as 
\begin{eqnarray}
    \epsilon (\gamma-1) \int_0^L q\left(\tilde{U}_x+\frac{D\tilde{x}}{Dt}\right)\mathrm{d}s=\gamma\int_0^L \left( \tilde{U}_y+\Omega\tilde{x}+\frac{D\tilde{y}}{Dt}\right)\mathrm{d}s+O(\epsilon^3, \epsilon^2\eta)
    \label{eq:Fy}.
\end{eqnarray}
Similarly, the torque balance, $\tilde{M}=0$, reads
\begin{eqnarray}
    \epsilon(\gamma-1)\int_0^L q\tilde{x}\left(\tilde{U}_x+\frac{D\tilde{x}}{Dt}\right)\mathrm{d}s+
    \int_0^L\tilde{y} \left(\tilde{U}_x+\frac{D\tilde{x}}{Dt}\right)\mathrm{d}s \nonumber \\
    =\gamma\int_0^L \tilde{x}\left( \tilde{U}_y+\Omega\tilde{x}+\frac{D\tilde{y}}{Dt}\right)\mathrm{d}s+O(\epsilon^3, \epsilon^2\eta)
    \label{eq:M}.
\end{eqnarray}
By similar order estimates for Eqs.~\eqref{eq:Fy}-\eqref{eq:M}, we find that in the leading order, only the right-most integral contributes in each equation. Hence, we obtain
\begin{eqnarray}
    L\tilde{U}_y^{(1)}+\frac{L^2}{2}\Omega^{(1)}+\epsilon\int_0^{L_0}\frac{\partial q}{\partial t}(s_0, t)\,\mathrm{d}s_0 &=&0,\\
    \frac{L^2}{2}\tilde{U}_y^{(1)}+\frac{L^3}{3}\Omega^{(1)}+\epsilon\int_0^{L_0}s_0\frac{\partial q}{\partial t}(s_0, t)\,\mathrm{d}s_0 &=&0,
\end{eqnarray}
yielding the first-order velocities,
\begin{eqnarray}
    \tilde{U}_y^{(1)}&=&-\frac{6\epsilon}{L^2}\int_0^{L_0}\left(\frac{2L}{3}-s_0\right)\frac{\partial q}{\partial t}\,\mathrm{d}s_0
    \label{eq:Uy1}, \\
    \Omega^{(1)}&=&\frac{12\epsilon}{L^3}\int_0^{L_0}\left(\frac{L}{2}-s_0\right)\frac{\partial q}{\partial t}\,\mathrm{d}s_0.
    \label{eq:Omega1}
\end{eqnarray}

\subsection{Second-order calculations}

We then proceed to the second-order velocities, by substituting the leading-order results into the force and torque balance equations [Eqs.~\eqref{eq:Fx}-\eqref{eq:M}].

We start with the tangential force balance. The second-order contributions are calculated as
\begin{equation}
    L\tilde{U}_x^{(2)}+\int_0^L\left(\tilde{W}_x-\Omega^{(1)}\tilde{y} \right)\,\mathrm{d}s+\epsilon(1-\gamma)\int_0^Lq\left(\tilde{U}_y^{(1)}+\Omega^{(1)}\tilde{x}+\frac{D\tilde{y}}{Dt} \right)\,\mathrm{d}s=0
    \label{eq:Fx2},
\end{equation}
where we introduced 
\begin{equation}
    \tilde{W}_x(s_0, t):=\tilde{U}^{(1)}_x+\frac{D \tilde{x}}{Dt}.
\end{equation}
From Eq.~\eqref{eq:Fx2}, we may readily obtain the second-order tangential velocity $\tilde{U}_x^{(2)}$.
Similarly, the normal force balance and torque balance equations are expanded up to the second order as
\begin{eqnarray}
    \epsilon (\gamma-1) \int_0^L q\tilde{W}_x\,\mathrm{d}s=\gamma\int_0^L \left( \tilde{U}^{(1)}_y+\Omega^{(1)}\tilde{x}+\frac{D\tilde{y}}{Dt}\right)\mathrm{d}s +\gamma\left( L\tilde{U}^{(2)}_y +\frac{L^2}{2}\Omega^{(2)}\right)
    \label{eq:Fy2},
\end{eqnarray}
and 
\begin{multline}
    \epsilon(\gamma-1)\int_0^L q\tilde{x}\tilde{W}_x\,\mathrm{d}s=
    -\int_0^L\tilde{y} \tilde{W}_x\,\mathrm{d}s
     +\gamma\int_0^L \tilde{x}\left( \tilde{U}^{(1)}_y+\Omega^{(1)}\tilde{x}+\frac{D\tilde{y}}{Dt}\right)\mathrm{d}s
    +\gamma\left( \frac{L^2}{2}\tilde{U}^{(2)}_y +\frac{L^3}{3}\Omega^{(2)}\right)
    \label{eq:M2},
\end{multline}
respectively. Hence, $\tilde{U}^{(2)}_y$ and $\Omega^{(2)}$ may be calculated by evaluating the integrals in Eqs.~\eqref{eq:Fy2}-\eqref{eq:M2} up to the second order.

\subsection{Third-order calculations}

To obtain the third-order progressive velocity, one needs to expand the position $\tilde{x}$ and $\tilde{y}$ and the vectors $\bm{n}$ and $\bm{t}$ up to the third order of small parameters, $\epsilon$ and $\eta$. The force balance equations in the $\tilde{x}$ direction [Eq.~\eqref{eq:Fx}] are therefore evaluated as 
\begin{multline}
    \tilde{U}^{(3)}_xL - \tilde{\Omega}^{(2)}\int_0^L\tilde{y}\,\mathrm{d}s+\int_0^L\left[\tilde{W}_x+\tilde{U}^{(2)}_y-\tilde{\Omega}^{(1)}\tilde{y}\right](1-\epsilon^2q^2)\,\mathrm{d}s+\epsilon^2\gamma\int_0^Lq^2\tilde{W}_x \,\mathrm{d}s\\
     + \epsilon(1-\gamma)\int_0^L\left[ \tilde{U}^{(1)}_y+\tilde{U}^{(2)}_y+\left(\tilde{\Omega}^{(1)}+\tilde{\Omega}^{(2)}\right)\tilde{x}+\frac{D\tilde{y}}{Dt}\right]\,\mathrm{d}s 
     =0, 
    \label{eq:Fx3}
\end{multline}
    by neglecting the fourth- and higher-order terms. Since the first- and second-order velocities are all calculated, by substituting these expressions into Eq.~\eqref{eq:Fx3} and evaluating the integrals, we may obtain the expression for the third-order tangential velocity $\tilde{U}^{(3)}_x$. 
    
\section{Bending wave with uniform compression}
\label{sec:uni_comp}

\subsection{Uniform compression}
\label{sec:asym_uni_comp}
As an illustrative example, we first consider locomotion with bending wave and uniform compression, given by
\begin{eqnarray}
    p(s_0, t)=\sin(\omega t)~\textrm{and}~q(s_0, t)=\sin(ks_0+\omega t+\phi)
    \label{eq:uni_comp_wave},
\end{eqnarray}
where $\phi\in[0, 2\pi)$ denotes the phase difference between the two modes.
For simplicity, we assume that the filament contains an integer number of waves, i.e., $kL_0=\pm 2\pi, \pm 4\pi, \pm 6\pi, \dots$. 

\begin{figure}[!t]
\begin{center}
\begin{overpic}[width=\linewidth]{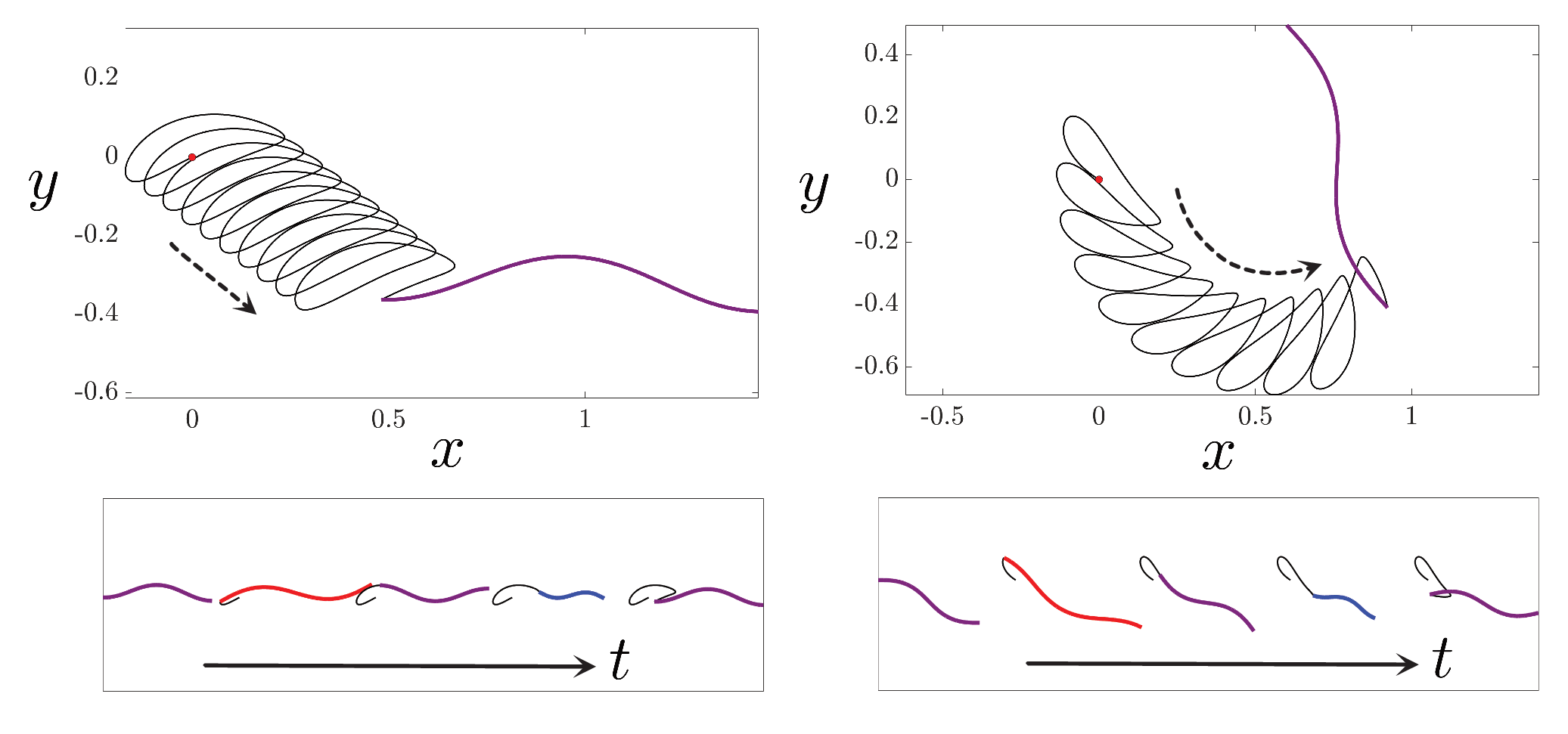}
\put(0, 43){(a)}
\put(50, 43){(b)}
\put(23, 42){$\phi=0$}
\put(68, 42){$\phi=\pi/2$}
\end{overpic}\\
\caption{ (Top) Sample trajectories of a swimmer under uniform compression with different phase shifts, (a) $\phi=0$ and (b) $\phi=\pi/2$. The parameters are  $\epsilon=\eta=0.4$, $\gamma=2$, $L_0=1, kL_0=2\pi$ and $\omega=1$. With the initial position $(0,0)$ (marked in a red circle) and the initial angle $\theta=0$, we drew the orbits of the leftmost end of the filament from $t=0$ to $t=10$. The configuration at $t=10$ is also shown. (Bottom) Time sequence of swimmer shape from $t=0$ to $t=1$ with the local extension visualised by the colours.}
\label{fig:traj_unicomp}
\end{center}
\end{figure}
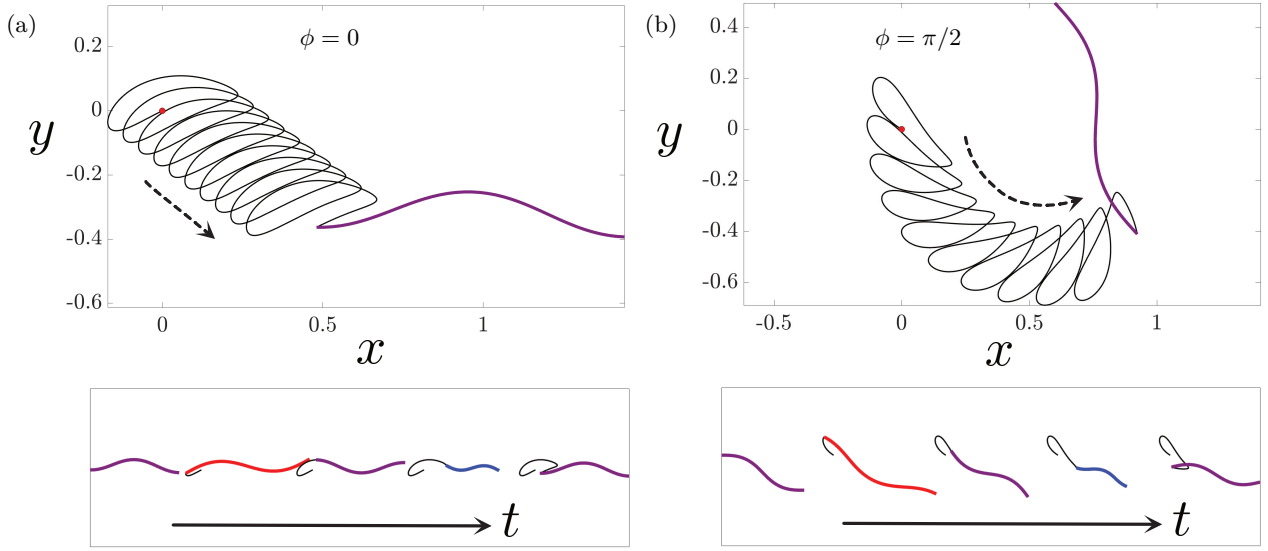

The first-order calculations provide
\begin{eqnarray}
    \tilde{U}_x^{(1)}&=&-\frac{\eta\omega L}{2}\cos\omega t \\
    \tilde{U}_y^{(1)}&=&\frac{\epsilon\omega}{k}\sin(\omega t+\phi) -\frac{6\epsilon\omega}{k^2L_0}\cos(\omega t+\phi)\\
    \Omega^{(1)}&=&\frac{12\epsilon\omega}{k^2L_0^2}\cos(\omega t+\phi),
\end{eqnarray}
which vanish after averaging over a time period. 

We then proceed to calculating the second-order contributions.
The tangential velocity at the second order may be calculated from Eq.~\eqref{eq:Fx2} as
\begin{equation}
    \tilde{U}^{(2)}_x=\frac{\epsilon^2\omega}{2k}\left[ (\gamma-1)+\frac{1}{2}\cos(2\omega t+2\phi)\right]
    -\frac{6\epsilon^2\omega}{k^3L_0^2}(\gamma-2)\left[1+\cos(2\omega t+2\phi)\right].
\end{equation}
By calculating the integrals and solving the linear problem at the second-order equations [Eqs.~\eqref{eq:Fy2}-\eqref{eq:M2}], we obtain
\begin{eqnarray}
    \tilde{U}_y^{(2)}&=&-\frac{\epsilon\eta\omega}{2 k}\left[ \left(\frac{\gamma-1}{\gamma}   \right)\cos\phi+\left(\frac{3\gamma-1}{\gamma}\right)\cos(2\omega t+\phi)\right] \\
    &&-\frac{3\epsilon\eta\omega}{k^2L_0^3}\left[ \left(\frac{2\gamma-3}{\gamma}\right)\sin\phi+\left(\frac{4\gamma-3}{\gamma}\right)\sin(2\omega t+\phi)\right] \\
    \Omega^{(2)}&=&\frac{18\epsilon\eta\omega}{k^2L_0^2}\left(\frac{\gamma-1}{\gamma}\right)\left[\sin\phi+\sin(2\omega t+\phi) \right]\label{eq:unicomp-Omega2}.
\end{eqnarray}
By taking the time average, we have
\begin{eqnarray}
 \langle\tilde{U}_x\rangle&=&\frac{\epsilon^2\omega}{2k}(\gamma-1)-\frac{6\epsilon^2\omega}{k^3L_0^2}(\gamma-2) \\
    \langle \tilde{U}_y\rangle&=&
    -\frac{\epsilon\eta\omega}{2k}\left(\frac{\gamma-1}{\gamma}\right)\cos\phi -\frac{3\epsilon\eta\omega}{k^2L_0}\left(\frac{2\gamma-3}{\gamma} \right)\sin\phi,
\end{eqnarray}
and for the angular velocity,
\begin{equation}
   \langle \Omega\rangle=
    \frac{18\epsilon\eta\omega}{k^2L_0^2}\left( \frac{\gamma-1}{\gamma}\right)\sin\phi
    \label{eq:UC_Omega}.
    \end{equation}

To obtain the expressions in the laboratory frame to the second order, one needs to take the non-commutative effects into account as in Eq.~\eqref{eq:magnus}, leading to
\begin{eqnarray}
    \langle U_x\rangle&=&\frac{\epsilon^2\omega}{2k}(\gamma-1)-\frac{6\epsilon^2\omega}{k^3L_0^2}(\gamma-1) \label{eq:UC-UxCor}\\
    \langle U_y\rangle&=&
    -\frac{\epsilon\eta\omega}{2k}\left(\frac{\gamma-1}{\gamma}\right)\cos\phi+\frac{9\epsilon\eta\omega}{k^2L_0}\left(\frac{\gamma-1}{\gamma} \right)\sin\phi, \label{eq:UC-UyCor}
\end{eqnarray}
while the rotational velocity is unchanged from Eq.\eqref{eq:UC_Omega}.
We clearly find no net locomotion with the isotropic drag ($\gamma=1$) for the uniform compression as in the general theory in \S\ref{sec:locomotion-with-isotropic-drag}.
 
The corrections by the Lie bracket appear in the second terms in Eqs.~\eqref{eq:UC-UxCor}-\eqref{eq:UC-UyCor}, which represent the finite-size effect of the body and vanish in the limit of $kL\rightarrow \infty$. In many mathematical models with bending motions including the Taylor sheet and a slender filament model usually assume one-dimensional motion, while this simplification is only valid for an infinitely-long body, because the instantaneous lateral and rotational motion, in general, does not vanish. This result highlights the importance of the motion non-commutativity for a finite-sized swimmer with lateral oscillation, the so-called yawing motion. 

Although the compression itself does not generate progressive velocity, the bending-compression coupling in the  $O(\epsilon\eta)$ terms generates both normal translation and rotation. This notably underlines the enhancement of manoeuvrability through body compression. In fact, by manipulating the phase shift, $\phi$, the swimmer is able to make a turn in both directions. 

When $\sin\phi=0$, as found from Eq.~\eqref{eq:UC_Omega}, no net rotation is generated.  
To analyse this no-net-rotation property in the large amplitude case, we numerically examine the swimmer dynamics, by discretising the slender swimmer by $N$ small segments with length $\ell_i$ ($i=1\dots N$), as is known as the $N$-link model \cite{moreau2018asymptotic}. We prescribe the segment length and the angle between the $i$-th and $i+1$-th segments, which we denote by $\alpha_i$ ($i=1\dots N-1$). The force balance and torque balances are then computed by summing up the contributions from each segment. These balance relations then  form linear equations for the translational and rotational velocities. In the numerical calculations below, we used $N=40$, which guarantees satisfying numerical accuracy for our purpose. 
Although the non-local hydrodynamic interactions are not negligible outside of the small-amplitude regime, we use the resistive drag relation as empirical modelling with a focus on the effects of the bending-compression coupling and the mechanical roles of drag anisotropy.
Sample trajectories of the swimmer are shown with its shape gait in Figure~\ref{fig:traj_unicomp}. The no-net-motion property still holds even outside of the small-amplitude regime, as seen in Figure~\ref{fig:traj_unicomp}(a). 

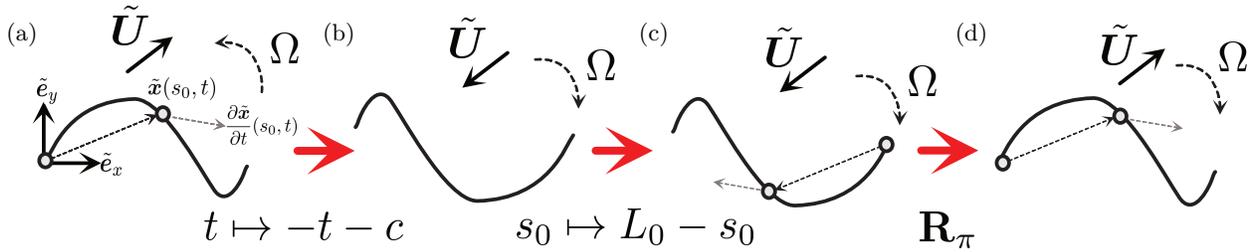
\begin{figure}[!t]
\begin{center}
\begin{overpic}[width=0.99\linewidth]{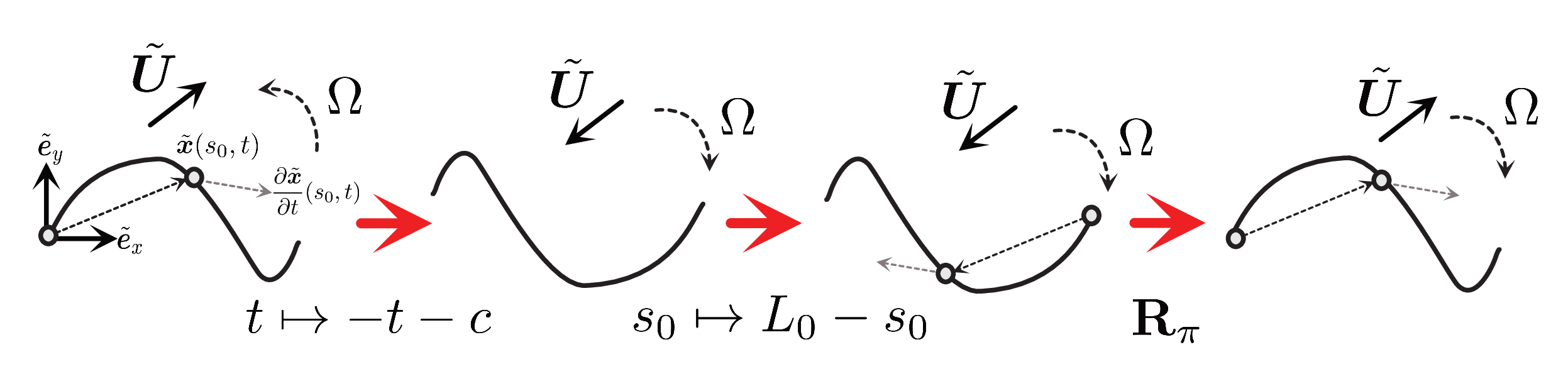}
\put(0, 18){(a)}
\put(25, 18){(b)}
\put(50, 18){(c)}
\put(75, 18){(d)}
\end{overpic}\\
\caption{Schematics for the symmetry arguments of the no-net-rotation. (a) The original shape in the body fixed frame with velocities $\tilde{\bm{U}}$ and $\Omega$. (b) The shape after the time reversal with a shift $c=kL_0/\omega-T/2$ (c) The head-to-tail inversion. (d) The $\pi$-rotation of the system.}
\label{fig:sym}
\end{center}
\end{figure}

In fact, this can be predicted by considerations on the symmetry of the shape gait.  Let us consider two linear changes of variables: time reversal with a phase shift $t\mapsto t'=-t+T/2+kL_0/\omega$, followed by head-to-tail inversion $s_0\mapsto s_0'=L_0-s_0$ [Figure \ref{fig:sym}(a)-(c)]. Then, the bending and compression functions are unchanged: $q(s_0', t')=q(s_0, t)$ and $p(s_0', t')=p(s_0, t)$ from the assumption on $kL_0$. Hence, the positions in the body-fixed frame also remain the same, after taking a $\pi$-rotation of the system as illustrated in Figure \ref{fig:sym}(d). With this transformation of $s_0$ and $t$, the time derivatives of $p$ and $q$ have an additional minus sign,
$\partial q'/\partial t' = -\partial q/\partial t$ and $\partial p'/\partial t' =-\partial p/\partial t$, yielding again the same shape velocity after the additional $\pi$-rotation of the system. Therefore, we
recover the same shape and shape deformation velocity, while the rotation direction is reversed after the sequences of transformations [Figure \ref{fig:sym}(d)]. To compensate for the uniqueness of the motion with given boundary conditions, the net rotation must vanish, while the instantaneous rotation is still allowed by the phase shift in the transformation of $t$.

With $\sin\phi\neq 0$, the above symmetry arguments are no longer available. Hence, as in the case of $\phi=\pi/2$ of Figure~\ref{fig:traj_unicomp}(b), net rotational motions are generated by the bending-compression coupling, allowing the swimmer to turn.

\subsection{Symmetric uniform compression}
\label{sec:sym_uni_comp}
We then consider a symmetrical situation, where the functions $p(s_0, t)$ and $q(s_0, t)$ are given by
\begin{eqnarray}
    p(s_0, t)=\sin(2\omega t)~\textrm{and}~q(s_0, t)=\sin(ks_0+\omega t+\phi)
    \label{eq:uni_comp_wave_sym},
\end{eqnarray}
to focus on progressive velocity. The assumption of an integer number of waves is again employed here.

By expanding the velocities up to the second order,
direct calculations then lead to the progressive velocity in the body-fixed coordinates, 
\begin{equation}
    \tilde{U}^{(2)}_x=\frac{\epsilon^2\omega}{2k}\left[ (\gamma-1)+\frac{1}{2}\cos(2\omega t+2\phi)\right]
    -\frac{6\epsilon^2\omega}{k^3L_0^2}(\gamma-2)\left[1+\cos(2\omega t+2\phi)\right],
\end{equation}

as well as normal and rotational velocities,
\begin{eqnarray}
    \tilde{U}_y^{(2)}&=&-\frac{\epsilon\eta\omega}{2 k}\left[ \left(\frac{3\gamma-2}{\gamma}   \right)\cos(\phi-\omega t)+\left(\frac{5\gamma-2}{\gamma}\right)\cos(3\omega t+\phi)\right] \\
    &&-\frac{3\epsilon\eta\omega}{k^2L_0}\left[ \left(\frac{5\gamma-6}{\gamma}\right)\sin(\phi-\omega t)+\left(\frac{7\gamma-6}{\gamma}\right)\sin(3\omega t+\phi)\right] \\
    \Omega^{(2)}&=&\frac{36\epsilon\eta\omega}{k^2L_0^2}\left(\frac{\gamma-1}{\gamma}\right)\left[\sin(\phi-\omega t)+\sin(3\omega t+\phi) \right].
\end{eqnarray}

By taking the time-average of the second-order velocities, we obtain 
\begin{eqnarray}
\langle\tilde{U}^{(2)}_x\rangle=\frac{\epsilon^2\omega}{2k}(\gamma-1)-\frac{6\epsilon^2\omega}{k^3L_0^2}(\gamma-2) 
\end{eqnarray}
and  $\langle \tilde{U}_y\rangle=\langle \Omega\rangle=0$ as expected by symmetry. After including the non-commutative effects between the normal translation and the rotation, we recover the same expression as in the asymmetrical case,
\begin{eqnarray}
    \langle U_x^{(2)}\rangle&=&\frac{\epsilon^2\omega}{2k}(\gamma-1)-\frac{6\epsilon^2\omega}{k^3L_0^2}(\gamma-1). 
\end{eqnarray}

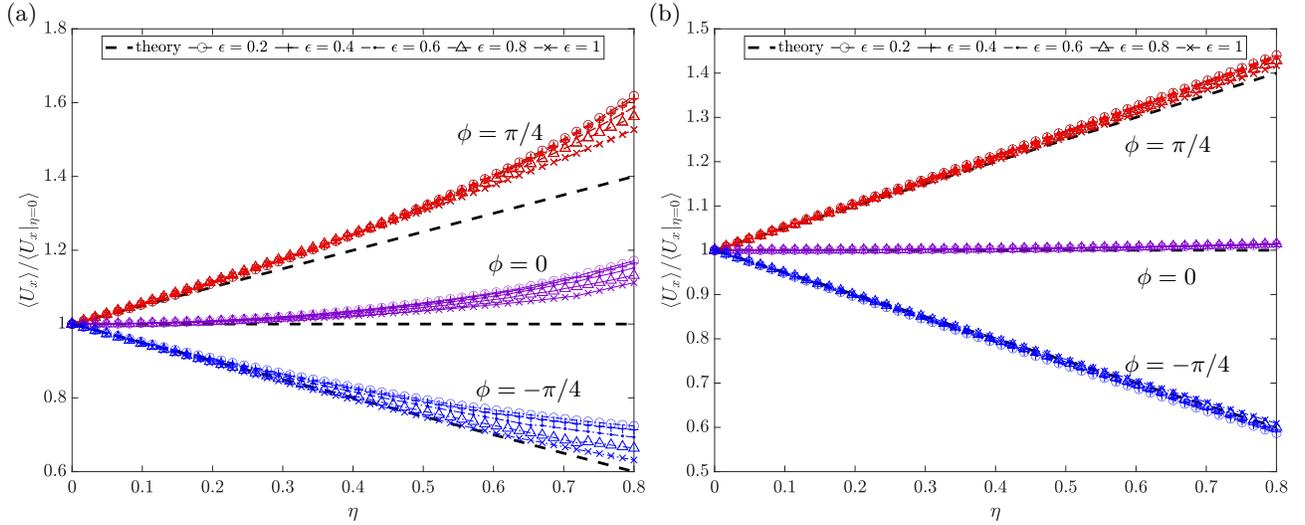
\begin{figure}[!t]
\begin{center}
\vspace{1em}
\begin{overpic}[width=0.49\linewidth]{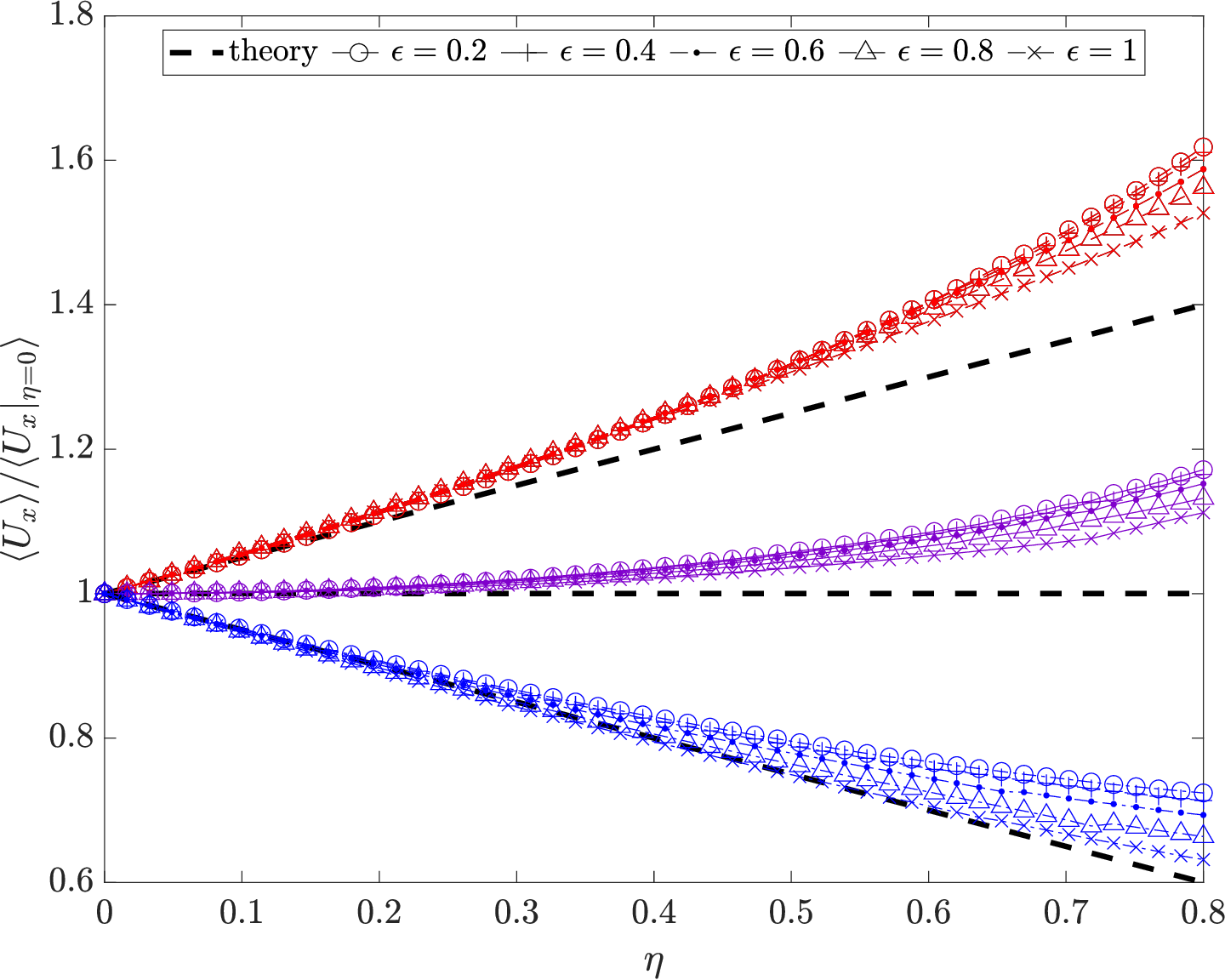}
\put(-2, 80){(a)}
\put(70, 61){$\phi=\pi/4$}
\put(75, 40){$\phi=0$}
\put(73, 20){$\phi=-\pi/4$}
\end{overpic}~
\begin{overpic}[width=0.49\linewidth]{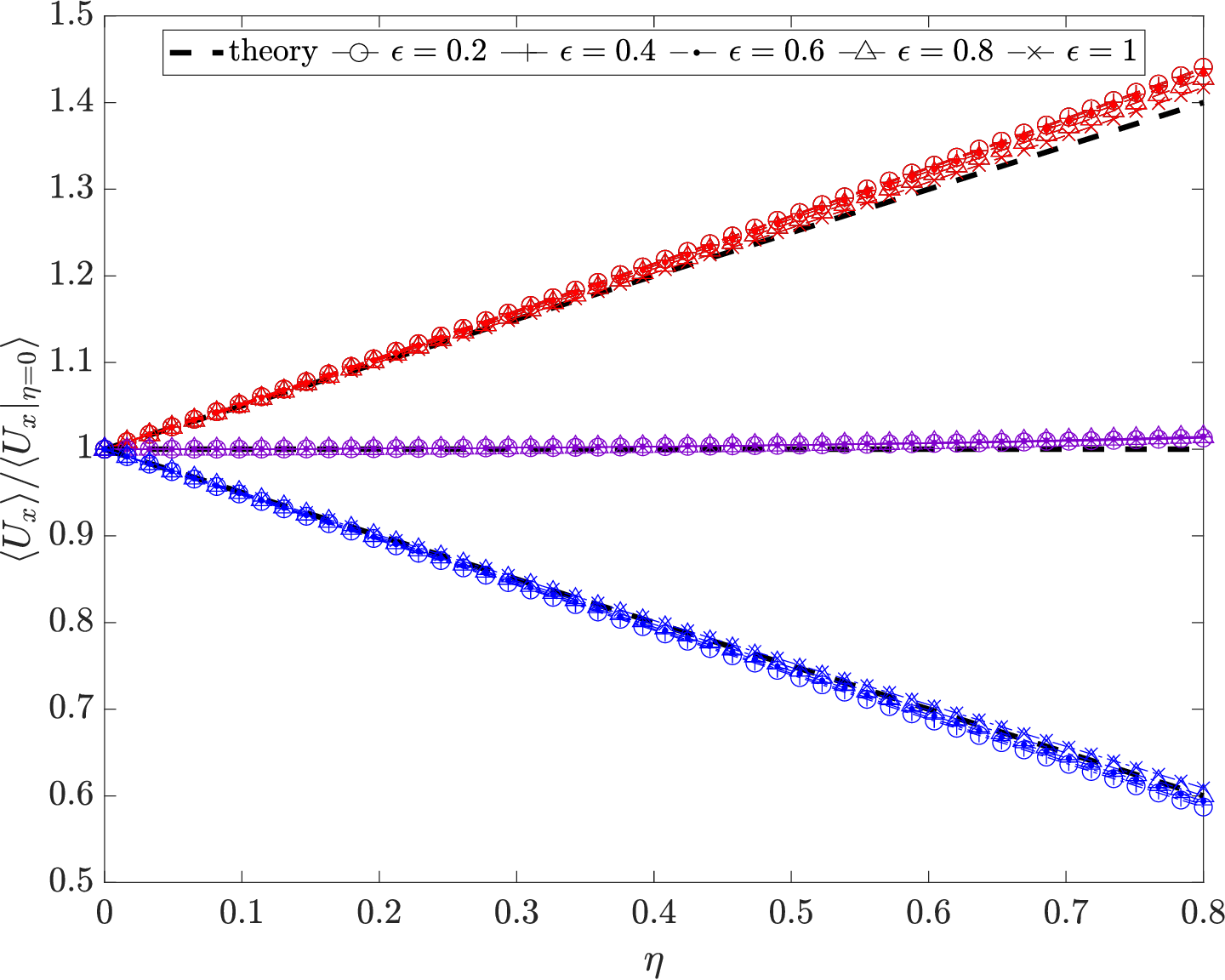}
\put(-2, 80){(b)}
\put(74, 59){$\phi=\pi/4$}
\put(76, 38){$\phi=0$}
\put(74, 24){$\phi=-\pi/4$}
\end{overpic}\\
\caption{Impacts of uniform compression on the swimmer velocity. The averaged velocity relative to a non-compressive bending swimmer. We used symmetrical uniform compression with  $\omega=2\pi$ and $\gamma=2$ for different values of $\eta$ and $\phi\in\{0, \pm\pi/4\}$. The wavenumber was used for $k=2\pi$ for (a) and $k=6\pi$ for (b).}
\label{fig:sym_uni_comp}
\end{center}
\end{figure}

Now, let us examine the third-order contribution for the tangential velocity. We need to expand $\tilde{x}(s_0, t)$ and $\tilde{y}(s_0, t)$ as well as the vectors $\bm{e}_{\|}(s_0, t)$ and $\bm{e}_{\perp}(s_0, t)$ up to the third orders of $\epsilon$ and $\eta$.
The third-order tangential velocity in the body-fixed frame is therefore obtained by substituting the first- and second-order velocities into Eq.~\eqref{eq:Fx3} and evaluating the integrals. Since the expression is lengthy, we only present the time-averaged velocity, $\langle \tilde{U}^{(3)}_x\rangle$, given by
\begin{equation}
    \langle \tilde{U}^{(3)}_x\rangle=\frac{\epsilon^2\eta\omega}{4k}(\gamma-1)\sin 2\phi-\frac{3\epsilon^2\eta\omega}{k^3 L_0^2}\left( \frac{(5\gamma-6)(\gamma-2)}{\gamma}\right)\sin2\phi.
\end{equation}

We then evaluate the Lie brackets in Eq.~\eqref{eq:magnus} from the matrix $\hat{\mathcal{A}}$ by employing the first- and second-order velocities as well as $\tilde{U}^{(3)}_x$. The averaged tangential velocity in the laboratory frame may be calculated up to the third order in the form, $\langle U_x\rangle=\langle U_x^{(2)}\rangle+\langle U_x^{(3)}\rangle$, with
\begin{equation}
    \langle U_x^{(3)}\rangle=\frac{\epsilon^2\eta\omega}{4k}(\gamma-1)\sin2\phi-\frac{3\epsilon^2\eta\omega}{k^3 L_0^2}\left( \frac{5\gamma-8}{\gamma}\right)(\gamma-1)\sin2\phi.
    \label{eq:Ux3Lab}
\end{equation}
Hence, thanks to the corrections coming from the non-commutativity of the rigid motion, we recover Case \ref{item : prop unif} of Proposition \ref{prop}.

The effect of uniform compression may be found at the third order with an increase or decrease of velocity depending on the phase difference $\phi$. We also numerically examine the effect of the uniform compression outside of the small-amplitude region, and plot the ratio of the averaged velocities with and without compression, $\langle U_{x}\rangle/\langle U_{x}|_{\eta= 0)}\rangle$ in Figure~\ref{fig:sym_uni_comp}. 
For small amplitude, we obtain that this ratio is expressed, up to the third order, by a linear function of $\eta$ as
\begin{eqnarray}
    \frac{\langle U_{x}\rangle}{\langle U_{x}|_{\eta= 0}\rangle}=1+\frac{\langle U^{(3)}_x\rangle}{\langle U^{(2)}_x\rangle}
    =\frac{\eta}{2}\left(1-\frac{15-24\gamma^{-1}}{k^2 L^2_0}\right)\left(1-\frac{6}{k^2L_0^2}\right)^{-1}.
\end{eqnarray}
As seen in the figure, the impact of the compression is linearly proportional to the compression size $\eta$, which well agrees with what the small-amplitude theory predicts. The quantitative agreements are more precise as the wavenumber $k$ increases [see Figure \ref{fig:sym_uni_comp}(b)]. 

Remarkably, the plots in Figure \ref{fig:sym_uni_comp} are only weakly affected by the size of $\epsilon$, although the small-amplitude theory cannot apply for a large $\epsilon$.


\section{Bending-compression wave}
\label{sec:comp_bend_wave}

\begin{figure}[!t]
\begin{center}
\begin{overpic}[width=\linewidth]{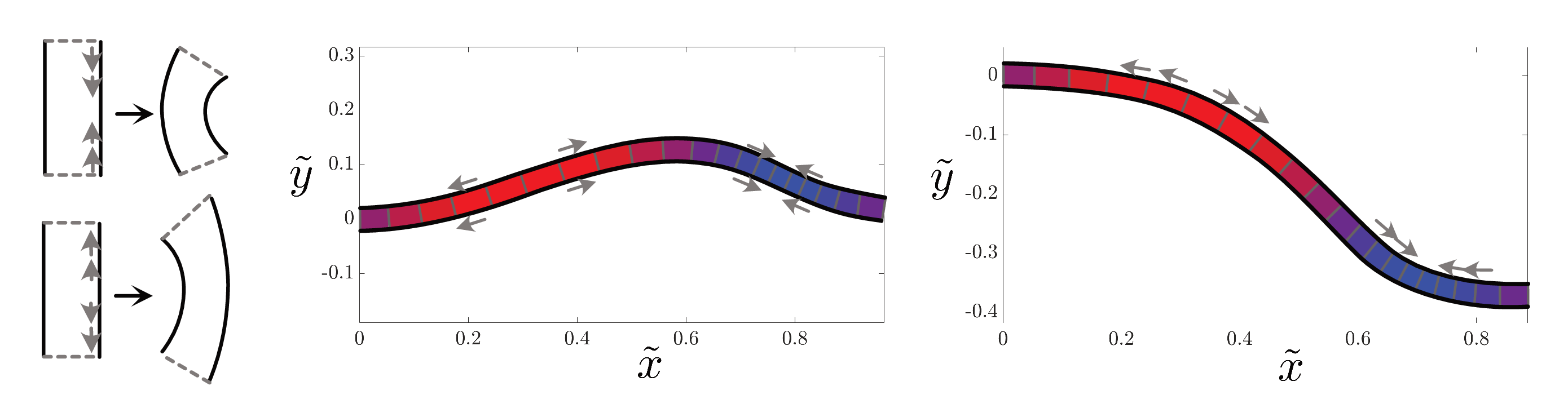}
\put(-2, 22){(a)}
\put(16, 22){(b)}
\put(59, 22){(c)}
\put(25, 18){$\phi=0$}
\put(65, 16){$\phi=\pi/2$}
\end{overpic}\\
\caption{(a) Schematic of contraction and extension of elastic body which exhibits bending and compression (shown by arrows).  (b, c) The configuration of the swimmer with bending-compression wave. The parameters are $\epsilon=\eta=0.4$, $L_0=1, kL_0=2\pi, \omega=1$ and phase shift is chosen as (b) $\phi=0$ and (c) $\phi=\pi/2$. We set the position as $(0,0)$ and the initial angle as $\theta=0$. The local extensibility of a slender body is shown by different colours from blue (compressed) to red (extended).}
\label{fig:comp_bend}
\end{center}
\end{figure}
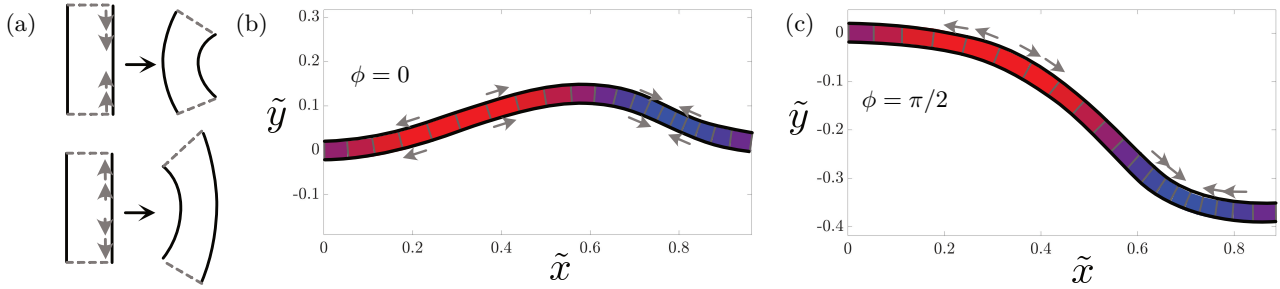

Motivated by muscular contraction, we then consider a compression-bending wave with a simple sinusoidal wave of $p$ and $q$, given by
\begin{eqnarray}
    p(s_0, t)=\sin(ks_0+\omega t)~,~~q(s_0, t)=\sin(ks_0+\omega t+\phi)
    \label{eq:comp_bend_wave_phi}.
\end{eqnarray}
As illustrated in Figure~\ref{fig:comp_bend}, local contraction of one side of a slender body yields both compression and bending.

For brevity, we again assume that the filament contains an integer number of waves, i.e., $kL_0=\pm 2\pi, \pm 4\pi, \pm 6\pi, \dots$. Also, to avoid displaying unnecessarily cumbersome expressions in the general case, we focus on two illustrative examples of $
\phi=0$  [Figure~\ref{fig:comp_bend}(b)] and $
\phi=\pi/2$ [Figure~\ref{fig:comp_bend}(c)].
This shape morphology is a counter-part example of the 'squeeze-me-bend-you' swimmer, with the total length $L$ of the body remaining constant in time due to the assumption of an integer wavenumber $k$.

\subsection{The $\phi=0$ case}
\label{sec:bend_comp_phi0}

\begin{figure}[!t]
\begin{center}
\begin{overpic}[width=\linewidth]{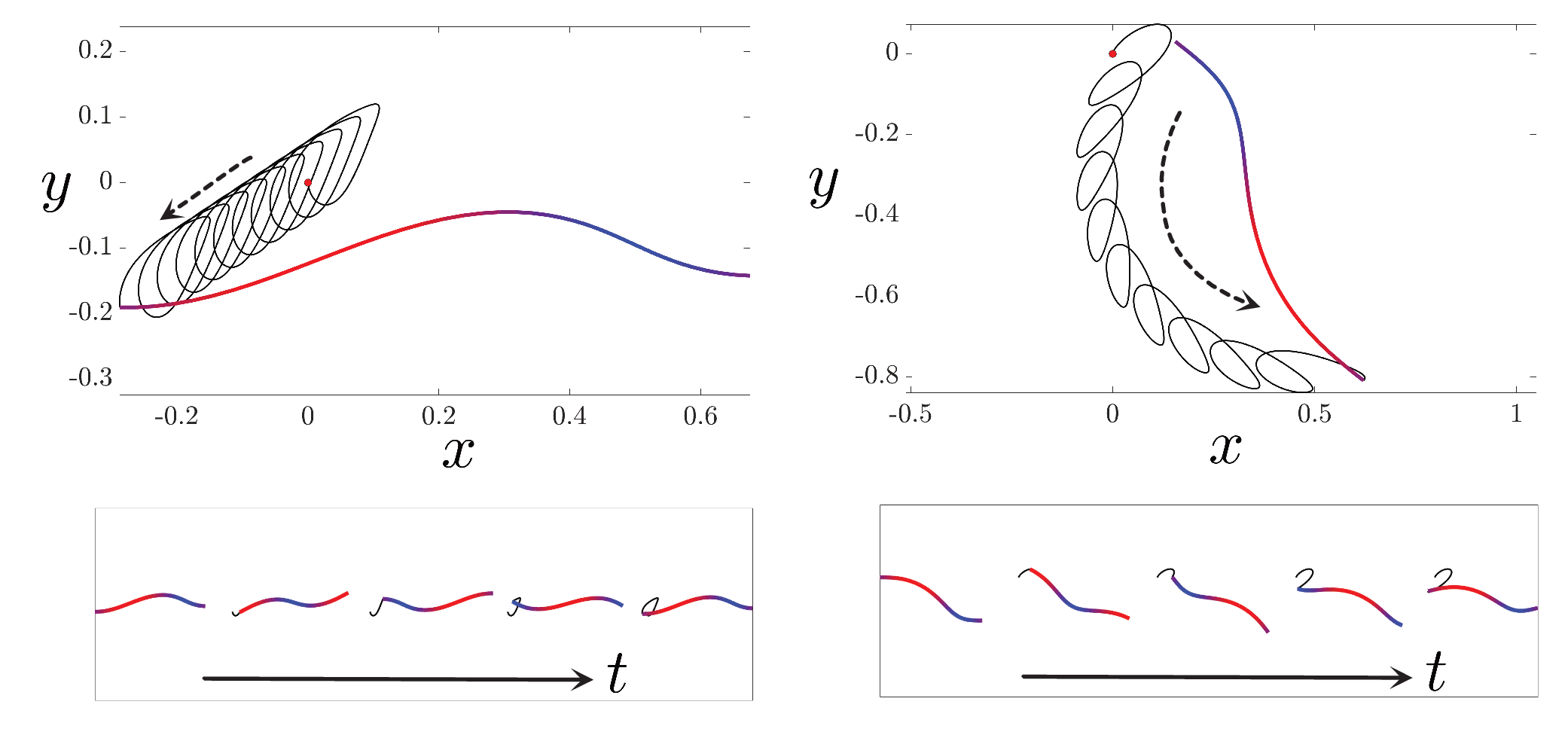}
\put(0, 43){(a)}
\put(50, 43){(b)}
\put(28, 41){$\phi=0$}
\put(80, 41){$\phi=\pi/2$}
\end{overpic}\\
\caption{(Top) Sample trajectories of a swimmer with bending-compression wave with (a) $\phi=0$ and $\phi=\pi/2$. The parameters are the same as in Figure \ref{fig:comp_bend}. With the initial position $(0,0)$ (marked in a red circle) and the initial angle $\theta=0$, we drew the orbits of the leftmost end of the filament from $t=0$ to $t=10$. The configuration at $t=10$ is also shown. (Bottom) Time sequence of swimmer shape from $t=1$ to $t=1$. The colour of the swimmer represents the local extension with the same colour as in Figure \ref{fig:comp_bend}.}
\label{fig:traj_comp_bend}
\end{center}
\end{figure}
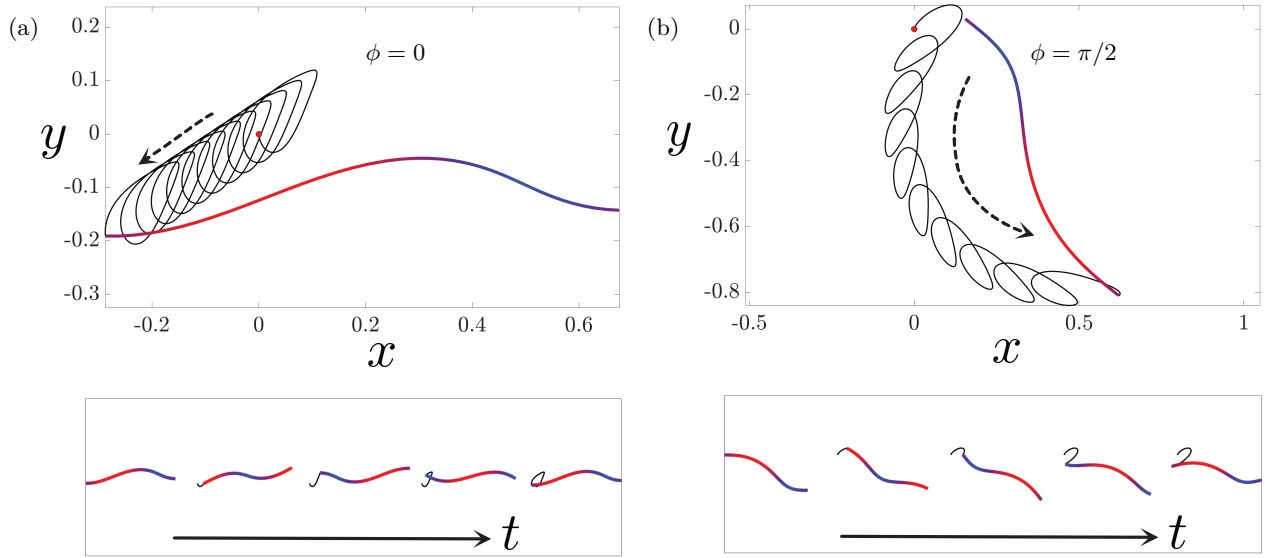

When $\phi=0$, the compression and bending, $p$ and $q$, are given by
\begin{eqnarray}
    p(s_0, t)=q(s_0, t)=\sin(ks_0+\omega t)
    \label{eq:comp_bend_wave}.
\end{eqnarray}
An example shape of the swimmer is shown in Figures.~\ref{fig:comp_bend}(a) and \ref{fig:traj_comp_bend}(a) and the local extension, $p(s_0, t)$, is visualised by the different colours from blue (compressed) to red (extended).

Using the results derived in the previous section [Eqs.~\eqref{eq:Ux1}, \eqref{eq:Uy1}-\eqref{eq:Omega1}], we readily obtain the leading-order velocities by direct calculations as
\begin{eqnarray}
       \tilde{U}^{(1)}_x&=&\frac{\eta\omega }{k}\sin\omega t
       \label{eq:CBW_Ux1}, \\
    \tilde{U}^{(1)}_y&=&\frac{\epsilon\omega}{k}\sin\omega t-\frac{6\epsilon\omega}{k^2 L_0}\cos\omega t ,
    \label{eq:CBW_Uy1}\\
    \Omega^{(1)}&=&\frac{12\epsilon\omega}{k^2 L_0^2}\cos\omega t . 
    \label{eq:CBW_Omega1}
\end{eqnarray}
The time average of these velocities all vanish at this order, $\langle \tilde{U}^{(1)}_x\rangle=\langle \tilde{U}^{(1)}_y\rangle=\langle \Omega^{(1)}\rangle=0$.

To examine the net locomotion, we proceed to the second-order calculations. The tangential velocity, $\tilde{U}^{(2)}_x$ may be obtained by substituting the leading-order expressions into Eq.~\eqref{eq:Fx2}. After some manipulations, we have
\begin{equation}
    \tilde{U}^{(2)}_x=\frac{\omega}{2k}\left[   \epsilon^2(\gamma-1)    -   \eta^2     
   + \frac{\epsilon^2}{2}   \cos(2 \omega t)\right]-\frac{6\omega\epsilon^2}{k^3L^3_0}   (\gamma-2)(1+\cos(2\omega t)).
\end{equation}
From the linear equations for $\tilde{U}^{(2)}_y$ and $\Omega^{(2)}$ [Eqs.~\eqref{eq:Fy2}-\eqref{eq:M2}], we may derive the second-order contributions:
\begin{eqnarray}
    \tilde{U}^{(2)}_y&=&-\frac{\epsilon\eta\omega}{2k}\left[\frac{1+\gamma\cos(2\omega t)}{\gamma}\right]-\frac{3\epsilon\eta\omega}{2k^2L_0}\left(\frac{\gamma+1}{\gamma}\right)\sin(2\omega t) \\
    \Omega^{(2)}&=&\frac{3\epsilon\eta\omega}{k^2L_0^2}\left(\frac{\gamma+1}{\gamma}\right)\sin(2\omega t).
\end{eqnarray}

To obtain the locomotion after a full deformation cycle from the instantaneous velocities, we consider the extra contribution from the non-commutative Lie algebra (see  Eq.~\eqref{eq:magnus}) in addition to the time average of the velocities in the body-fixed frame, which gives
\begin{eqnarray}
    \langle U^{(2)}_x\rangle&=&
    \frac{\omega}{2k}\left[(\gamma-1)\epsilon^2-\eta^2\right]-(\gamma-1)\frac{6\omega}{k^3L_0^2}\epsilon^2, \\
    \langle U^{(2)}_y\rangle&=&
    -\frac{\epsilon\eta\omega}{2\gamma k}+\frac{6\epsilon\eta\omega}{k^3L_0^2},
\end{eqnarray}
and the average angular velocity is simply the time average in the body-fixed frame, $\langle \Omega\rangle =0$.

The tangential velocity $\langle U^{(2)}_x\rangle$ contains an additional compression term of order $O(\eta^2)$, which remains for a purely compression/extension wave without bending. Note that the direction of swimming from the compression wave is opposite to the locomotion from the bending wave.

The normal velocity $\langle U^{(2)}_y\rangle$ does not vanish. This notable effect of the drift velocity is generated by the bending-compression coupling, seen as the $O(\epsilon\eta)$-term. 
The rotational motion, however, vanishes after taking the time average.
To examine the dynamics outside of the small-amplitude theory, we numerically solve the system and plot an example trajectory in Figure~\ref{fig:traj_comp_bend}(a). As seen in the figure, 
no-net-rotation still holds for a large amplitude and this property is derived by the symmetry arguments as in the uniform compression case, where we consider the $\pi$-rotation of the system after the time reversal with a phase shift and the head-to-tail inversion, $t\mapsto t'=-t+T/2-kL_0/\omega$ and $s_0\mapsto s_0'=L_0-s_0$ (see also figure~\ref{fig:sym}). In this case, the symmetry arguments follow for any wavenumber $k$.


\subsection{The $\phi=\pi/2$ case}
\label{sec:bend_comp_phipi2}
We then consider the situation with $\phi=\pi/2$, where the contraction of one side of the body travels down, and 
an example shape is shown in Figs.~\ref{fig:comp_bend}(b) and \ref{fig:traj_comp_bend}(b) with different colours illustrating the compressed (blue) and extended (red) regions. 

We start with the small-amplitude regime, where we follow similar calculations as in the previous sections. 
By executing the calculations up to the second order and calculating the Lie brackets in Eq.~\eqref{eq:magnus}, we may derive the expressions of the velocities in the laboratory frame as
\begin{eqnarray}
    \langle U^{(2)}_x\rangle&=&
    \frac{\omega}{2k}\left[(\gamma-1)\epsilon^2-\eta^2\right]-(\gamma-1)\frac{6\omega}{k^3L_0^2}\epsilon^2, \\
    \langle U^{(2)}_y\rangle&=&
    -\frac{3\epsilon\eta\omega}{k^2L_0}\left(\frac{\gamma+1}{\gamma}\right) \\
    \langle \Omega\rangle&=&
    \frac{6\epsilon\eta\omega}{k^2L_0^2}\left(\frac{\gamma+1}{\gamma}\right).
\end{eqnarray}

The velocity $\langle U_x\rangle$ has the same expression as that of the  $\phi=0$ case, while the normal and rotational velocities are qualitatively different. In particular, net rotational motion is generated. An example trajectory with its shape gait is shown in Figure~\ref{fig:comp_bend}(b).

\subsection{Symmetric compression-bending wave}
\label{sec:sym_bend_comp}

Muscular contraction is typically generated symmetrically to avoid drifting and turning motions. In this subsection, therefore, we consider a symmetric bending-compression wave, given by
\begin{eqnarray}
    p(s_0, t)=\sin(ks_0+2\omega t)~\textrm{and}~q(s_0, t)=\sin(ks_0+\omega t+\phi)
    \label{eq:comp_bend_wave_sym}.
\end{eqnarray}
For brevity, we again assume that the filament contains an integer number of waves, i.e., $kL=\pm 2\pi, \pm 4\pi, \pm 6\pi, \dots$.

We first consider the case with $\phi=0$. The first-order calculations provide
\begin{eqnarray}
    \tilde{U}_x^{(1)}&=&\frac{2\eta\omega}{k}\sin(2\omega t) \\
    \tilde{U}_y^{(1)}&=&\frac{\epsilon\omega}{k}\sin(\omega t) -\frac{6\epsilon\omega}{k^2L_0}\cos(\omega t)\\
    \Omega^{(1)}&=&\frac{12\epsilon\omega}{k^2L_0^2}\cos(\omega t),
\end{eqnarray}
which yields no net motion or rotation. 
Up to the second order, we may obtain explicit forms by estimating the integrals in Eq.~\eqref{eq:Fx2} as
\begin{eqnarray}
    \tilde{U}^{(2)}_x&=&-\frac{\eta^2\omega}{k}+\frac{\epsilon^2\omega}{4k}\left[ 3-2\gamma-2\cos^2\omega t \right]-\frac{12\epsilon^2\omega}{k^3L_0^2}(\gamma-2)\cos^2\omega t \label{eq:symCBW-Ux2} \\
    \tilde{U}^{(2)}_y&=& \frac{\epsilon \eta\omega}{4 k}\left[ 2\left(\frac{\gamma-2}{\gamma}\right)\cos\omega t-3\cos(3\omega t)\right]\nonumber \\
    &&+\frac{\epsilon\eta\omega}{4 k^2 L_0}\left[ 12\left(\frac{\gamma+1}{\gamma}\right)\sin\omega t-3\left(\frac{\gamma+4}{\gamma}\right)\sin(3\omega t)\right]\\
    \Omega^{(2)}&=&\frac{\epsilon\eta\omega}{2}\sin\omega t+\frac{\epsilon\eta\omega}{2 k^2 L_0^2}\left[6\left(\frac{\gamma+3}{\gamma}\right)\cos\omega t -3\left(\frac{3\gamma+4}{\gamma}\right)\right]\sin\omega t
    \label{eq:symCBW-Omega2}.
\end{eqnarray}
By taking time-average, we derive
\begin{eqnarray}
    \langle\tilde{U}^{(2)}_x\rangle&=&\frac{\omega}{2k}\left[ (\gamma-1)\epsilon^2-2\eta^2\right]-\frac{6\epsilon^2\omega}{k^3L_0^2}(\gamma-2),
\end{eqnarray}
and zero normal and angular velocities,
\begin{equation}
    \langle \tilde{U}^{(2)}_y\rangle=0~,~~
    \langle \Omega^{(2)}\rangle=0,  
\end{equation}
as expected from the symmetry.

After the non-commutative algebra, the averaged velocity for the tangential velocity in the reference frame is calculated as
\begin{eqnarray}
    \langle U_x^{(2)}\rangle&=&\frac{\omega}{2k}\left[ (\gamma-1)\epsilon^2-2\eta^2\right]-\frac{6\epsilon^2\omega}{k^3L_0^2}(\gamma-1).
    \label{eq:CBWsym_Ux}
\end{eqnarray}

We then proceed to the third-order contributions for the progressive velocity, using Eq.~\eqref{eq:Fx3} and the second-order expressions [Eqs.\eqref{eq:symCBW-Ux2}-{\eqref{eq:symCBW-Omega2}]. Direct calculations lead to the time-average velocity,
\begin{equation}
    \langle \tilde{U}^{(3)}_x\rangle=\frac{3\epsilon^2\eta\omega}{2k^2L^2_0},
    \label{eq:symBCW-Ux3}
\end{equation}
which is, however, found to be cancelled out in the laboratory frame after calculating the Lie brackets such that
\begin{equation}
    \langle U^{(3)}_x\rangle = 0.
\end{equation}
The zero velocity contributions at the third order are similar to that of the classical Taylor sheet \cite{taylor1951analysis}, but are different from the uniform compression case.

\begin{figure}[!t]
\begin{center}
\begin{overpic}[width=10cm]{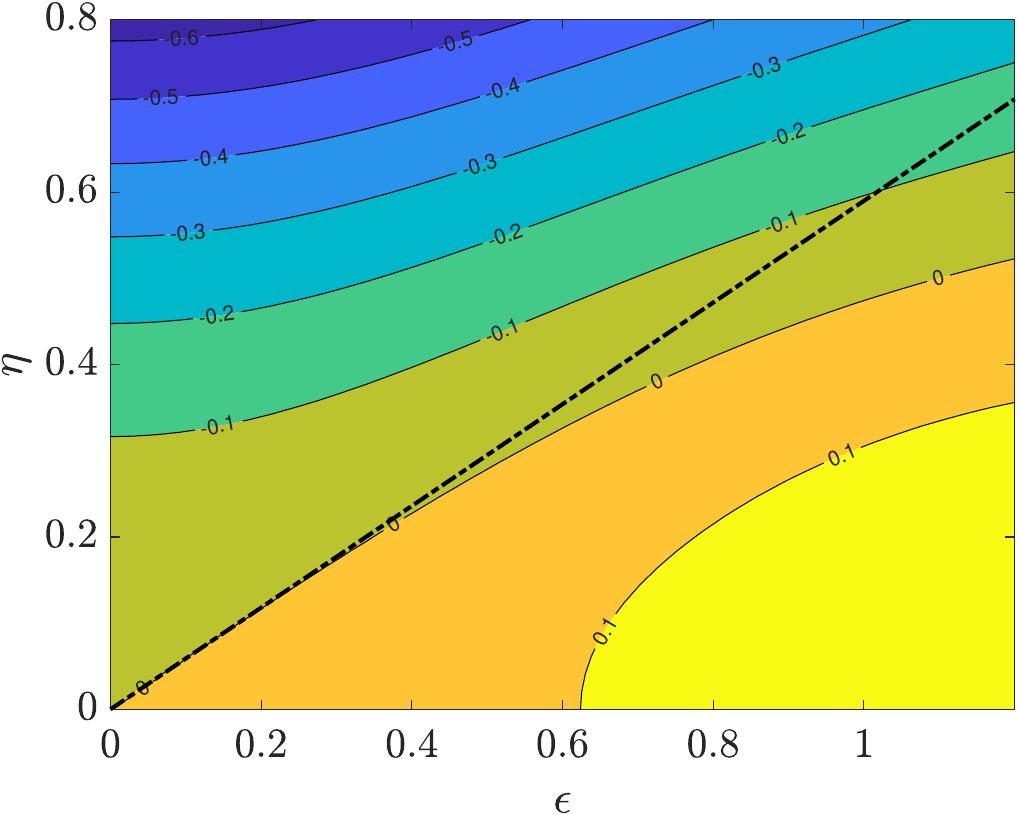}
\end{overpic}\\
\caption{Numerical simulation of averaged swimming velocity $\langle U_x \rangle$ (in the reference frame) of the swimmer driven by symmetrical compression wave with $L_0=1$, $kL=2\pi$, and $\omega=2\pi$ for different values of $\epsilon$ and $\eta$. The broken line is a null-curve of the velocity predicted by the small-amplitude theory up to the third order, which is in  excellent agreement when $\epsilon \lesssim 0.4$.}
\label{fig:sym_comp_wave}
\end{center}
\end{figure}

The case with $\phi=\pi/2$ can also be calculated in a similar manner.  
By executing the second-order calculations, we found that the expression of the averaged progressive velocity, $\langle U_x\rangle$, is identical to that obtained in the $\phi=0$ case [Eqs.\eqref{eq:CBWsym_Ux}], together with zero normal and rotational velocities due to the symmetry.
At the third order, the progressive velocity has a similar form as in  Eq.~\eqref{eq:symBCW-Ux3}:
\begin{equation}
    \langle \tilde{U}^{(3)}_x\rangle=-\frac{3\epsilon^2\eta\omega}{2k^2L^2_0},
    \label{eq:symBCW-Ux3}
\end{equation}
and, after including the non-commutative effects from the Lie bracket, this contribution vanishes again: $\langle U^{(3)}_x\rangle = 0$.

Hence, both when $\phi=0$ and $\phi=\pi/2$, according to the analysis above [Eq.~\eqref{eq:CBWsym_Ux}], net locomotion disappears when
\begin{equation}
\left|\frac{\eta}{\epsilon}\right|=
\sqrt{\left(1-\frac{12}{k^2L_0^2}\right)\frac{\gamma-1}{2}} 
        \label{eq:CBW_Ux_sym}.
\end{equation}

To test these theoretical results, 
we numerically compute the averaged swimming velocity for a finite amplitude with different $\epsilon$ and $\eta$. The results are shown in Figure~\ref{fig:sym_comp_wave}, together with the theoretical prediction of the null curve by the small-amplitude theory [Eq.~\eqref{eq:CBW_Ux_sym}]. As seen in the figure, the theoretical prediction is in excellent agreement when $\epsilon\lesssim 0.4$.


\begin{figure}[!t]
\begin{center}
\begin{overpic}[width=0.49\linewidth]{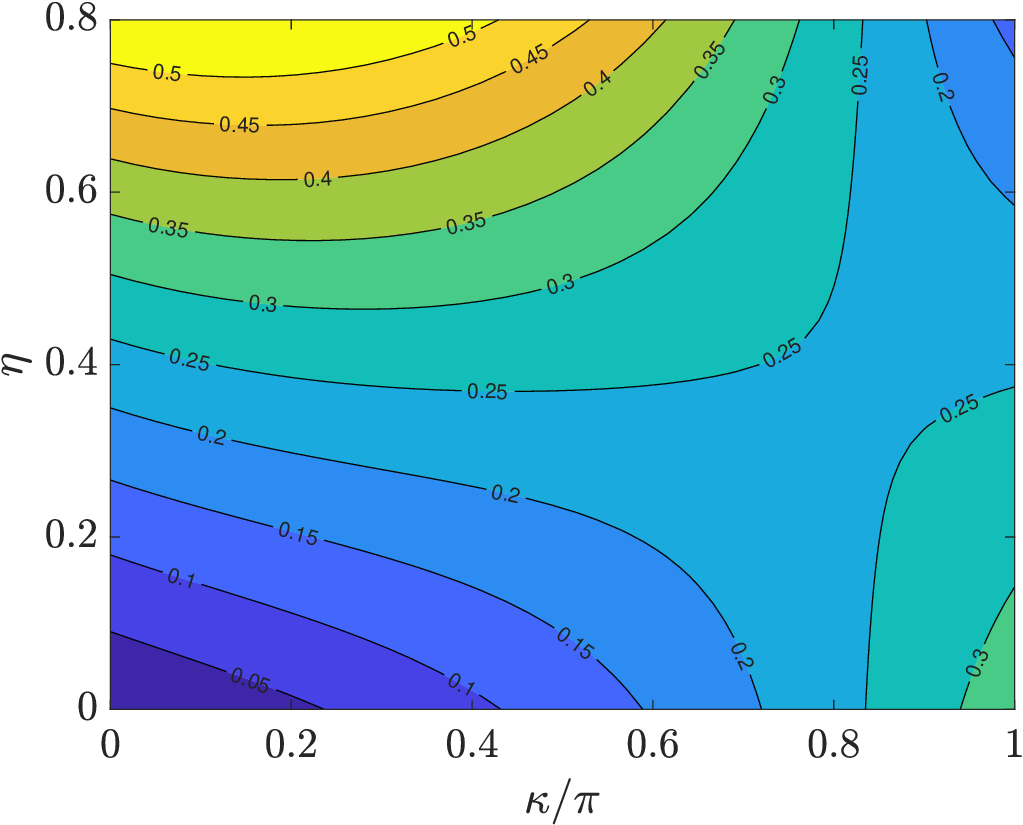}
\put(0, 72){(a)}
\end{overpic}
\begin{overpic}[width=0.49\linewidth]{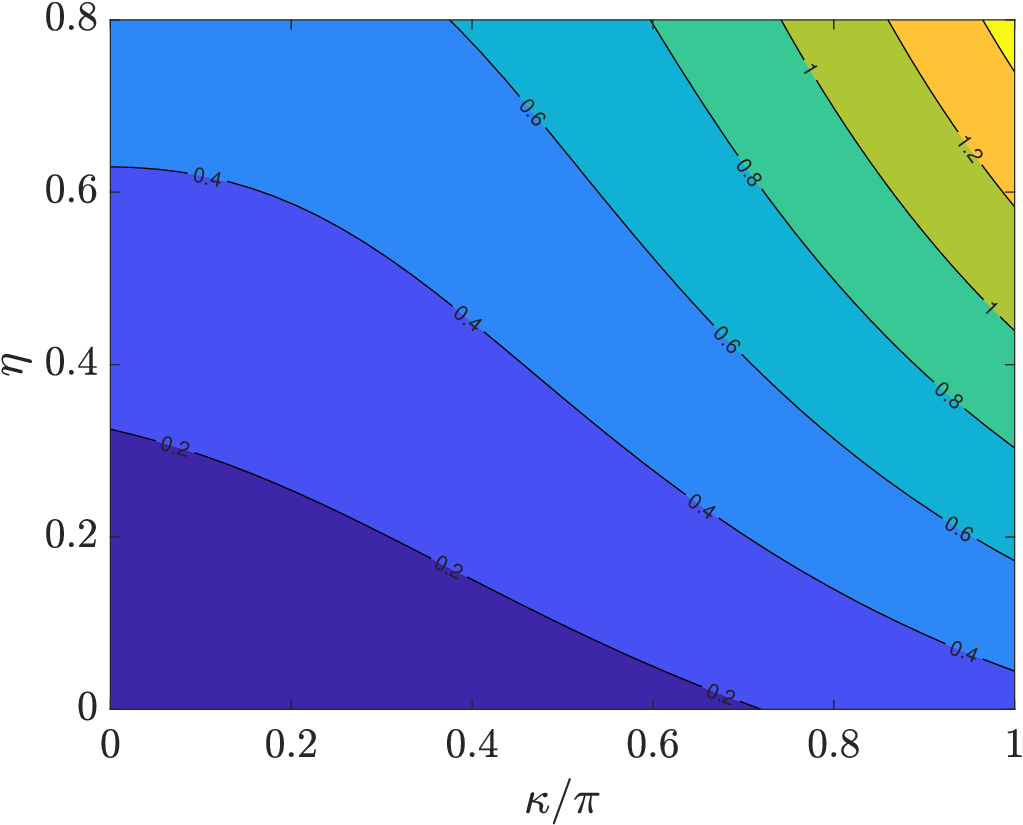}
\put(0, 72){(b)}
\end{overpic}
\caption{Numerical simulation of averaged angular velocity $\langle \Omega \rangle$ (in the reference frame) of the swimmer driven by (a) uniform compression wave and (b) bending-compression wave with $L_0=1$, $kL=2\pi$, and $\omega=2\pi$, $\epsilon=0.4$, $\phi=0.5\pi$ for different values of $\kappa\in[0, \pi]$ and $\eta\in[0, 0.8]$.}
\label{fig:kappa_eta}
\end{center}
\end{figure}

\section{Manoeuvrability by bending-compression bending}
\label{sec:manoeuvrability}

In the previous sections, we have seen that a phase difference between the bending and compression modes may generate swimmer rotation. 
To further qualitatively examine the manoeuvrability of extensible swimmers, in this section, we compare the turning dynamics by bending-compression coupling with those of inextensible slender swimmers.

Inextensible microswimmers with cilia and flagella control their orientation by laterally asymmetric beating, as seen in chemotactic sperm cells \cite{shiba2008ca2+}. To represent the asymmetric waveform, we consider a non-zero mean curvature in the bending angle as
\begin{equation}
    \tilde{\theta}(s_0, t)=\epsilon q(s_0, t)+\kappa s_0,
\end{equation}
where a constant $\kappa$ corresponds to the mean curvature in the absence of compression. 

We first perform perturbation analysis with the small-amplitude theory in \S\ref{sec:small_amp}, by assuming $\epsilon ,\eta, \kappa\ll1$,  $\eta=O(\epsilon)$, and $\kappa=O(\epsilon)$. Following the previous analyses in \S\ref{sec:uni_comp} and \S\ref{sec:comp_bend_wave}, we consider functions of $p(s_0, t)$ and $q(s_0, t)$ for uniform compression [Eq.~\eqref{eq:uni_comp_wave}] and compression-bending wave [Eq.~\eqref{eq:comp_bend_wave_phi}]. With the assumptions of the integer number of waves, $kL_0=\pm 2\pi, \pm4\pi, \dots$, the difference of the tangent angle is simply $\tilde{\theta}(L_0, t)-\tilde{\theta}(0, t)=\kappa L_0$.

We then redo the perturbation calculations of Sections \S\ref{sec:uni_comp} and \S\ref{sec:comp_bend_wave},  including the non-zero mean curvature. 
Resulting expressions reveal that in both the uniform compression and compression-bending wave cases, the time-averaged translational and rotational velocities are unchanged up to the second order of expansion, regardless of the non-zero mean curvature. 
In particular, the lateral and rotational velocities are $\langle U_y\rangle=O(\epsilon\eta)$ and $\langle \Omega\rangle=O(\epsilon\eta)$, respectively, and the effects of $\kappa$ only appear in the higher-order averaged velocities. Moreover, we note that $O(\epsilon\kappa)$ terms appear in the instantaneous velocities but all vanish by taking the time average. This quantitatively demonstrates the importance of the compression-driven turning in terms of swimmer manoeuvrability compared to asymmetric inextensible bending.

We then examine the swimmer manoeuvrability for the finite-amplitude case by numerically evaluating the time-averaged rotational velocity for different values of $\eta$ and $\kappa$, while the other parameters are fixed. In Figure~\ref{fig:kappa_eta}, we show an example of the contour plot of $\langle \Omega \rangle$ for the uniform compression  [Figure~\ref{fig:kappa_eta}(a)] and the compression-bending [Figure~\ref{fig:kappa_eta}(b)] cases. 
In both panels, the contour is rather horizontal in the small $\kappa$ regime, suggesting that asymmetric beating with non-zero mean curvature could be less efficient than compression-bending coupling to generate net rotation.


\section{Conclusions}
\label{sec:conc}

\begin{table}
  \begin{center}
\def~{\hphantom{0}}
  \begin{tabular}{c c  c  c c  c  c}
      & \multicolumn{3}{c}{ Uniform compression }  &  \multicolumn{3}{c}{Bending-compression wave }  \\  \hline
      Case & $\phi = 0, \pi$ & $\phi \neq 0, \pi$ & symmetric &  $\phi = 0$ &  $\phi = \pi/2 $ & symmetric \\ 
      Section & $\quad$ \S\ref{sec:asym_uni_comp} $\quad$ & $\quad$ \S\ref{sec:asym_uni_comp} $\quad$ & $\quad$ \S\ref{sec:sym_uni_comp} $\quad$ & $\quad$ \S\ref{sec:bend_comp_phi0} $\quad$ & $\quad$ \S\ref{sec:bend_comp_phipi2} $\quad$ & $\quad$ \S\ref{sec:sym_bend_comp} $\quad$ \\  \hline
      $\langle U_x^{(2)} \rangle$ & $\epsilon^2$ & $\epsilon^2$ & $\epsilon^2$ & $\epsilon^2 + \eta^2$ & $\epsilon^2 + \eta^2$ & $\epsilon^2 + \eta^2$ \\ 
      $\langle U_x^{(3)} \rangle$ & - & - & $\epsilon^2 \eta$ & - & - & 0 \\ 
      $\langle U_y^{(2)} \rangle$ & $\epsilon \eta$ & $\epsilon \eta$ & 0 & $\epsilon \eta$ & $\epsilon \eta$ & 0 \\ 
      $\langle \Omega^{(2)} \rangle$ & 0 & $\epsilon \eta$ & 0 & 0 & $\epsilon \eta$ & 0
  \end{tabular}
  \caption{Summary of the orders of magnitude appearing in the small amplitude analysis. Amplitudes of bending and compression are denoted as $\eta$ and $\epsilon$, respectively.}
  \label{tab:summary}
  \end{center}
\end{table}

 We have theoretically investigated the impact of compression or extension on the bending motion of slender objects at low Reynolds number, with a particular focus on the bending-compression coupling effects, motivated by some biological and artificial microswimmers that exhibit a large amount of compression and extension. 

 We first revisited the effect of body extensibility on swimming with isotropic drag, first discussed by \cite{pak2011extensibility}, refining the arguments to prove that no net motion is possible by uniform compression with an isotropic drag (Proposition \ref{prop}).
 Since the compression-bending coupling, in general, generates an instantaneous rotation or yawing motion, we employed the gauge theoretic formulation and the Lie bracket to deal with the rotation-translational coupling.

 We then introduced minimal theoretical models with two degrees of freedom; one from bending and the other from compression/extension. The first model, which we referred to as the compressive scallop swimmer, undergoes uniform compression. The second minimal model that we introduced as 'squeeze-me-bend-you' swimmer has a constant total length but the position of the hinge can vary over time. We computed the curvature field that characterises the locomotion and the results illustrated several deformation modes to generate translation and rotation, highlighting the enhancement of manoeuvrability and complexity of emergent dynamics through the compression during swimming.

 We then examined a general slender body, with systematic perturbation analyses within the small-amplitude regime both for bending angle amplitude $\epsilon$ and compression amplitude $\eta$ with $\epsilon, \eta\ll 1$.
 As counterpart examples to the minimal models, we theoretically and numerically examined in detail the uniform compression as well as compression-bending travelling wave, both being motivated by biological swimmers and relevant also in robotics.

 The results are summarised in Table~\ref{tab:summary}.
 The analysis shows that bending-compression coupling triggers transverse drift and turning behaviours, at the order of $O(\epsilon\eta)$. With the coupling effects, the swimmer can make a turn, although the emergent behaviours depend on the phase shift between the bending and compression. 
 This result highlights the importance of the shape gait symmetry, and we have found that net locomotion is only generated by the breakdown of symmetry.
 
 To focus on the progressive velocity of the swimmer, we then considered uniform compression with a double frequency. Detailed perturbation analysis found that the bending-compression coupling appears as the increase or decrease of swimming speed at the order of $O(\epsilon^2\eta)$ in the uniform compression case, while the higher-order contributions vanish for the bending-compression wave case. In both cases, the theoretical prediction well explained the numerical simulations, illustrating the usefulness of the systematic perturbation analysis incorporating the non-commutative effects of translation and rotation through the higher-order Lie brackets. 
  Such a non-commutative effect has long been neglected in theoretical calculations of microswimmer models such as the Taylor sheet and slender filaments, for example by assuming an infinite body length and one-dimensional locomotion. In contrast, we have explicitly derived the finite-size effect of a slender bending filament at the Stokesian regime. 
  The motion non-commutativity is inevitable for finite-sized bending swimmers which possess in general instantaneous lateral and rotational velocities. Such yawing effects in microswimmers are usually considered rapid time-scale dynamics and often naively averaged out. However, if the swimmer dynamics are coupled with outer environments, the long-time behaviour is no longer a simple average due to non-linear interactions. This has been recently studied through multiple scale analysis for coupling with external flows, boundaries, neighbours and spatial viscosity variations \cite{walker2022effects, walker2022emergent, walker2023systematic, kanazawa2024locomotion}.
  
Our results on bending-compression coupling unveiled emergent drifting and rotation dynamics as well as modulations of progressive velocity by non-linear interactions, which could be beneficial for higher manoeuvrability of locomotion. Also, these findings will be useful for better understanding of biological functionality in microswimmers with compression and extensibility \cite{yanase2018neck, wan2019ciliate, cammann2025form} as well as for designing the soft artificial microrobots using the compression/extension degree of freedom such as hydrogels \cite{nocentini2018optically}. Our study sheds light on compression as an important degree of freedom in locomotion at low Reynolds number and more broadly in a dissipative medium, including slithering and crawling animals on the ground. Also, the theoretical methodology based on the geometrical formulation of microswimming will be widely applicable to yawing behaviours and non-commutative translation-rotation coupling in microswimming.

\section*{Acknowledgments}
K.I. acknowledges the Japan Society for the Promotion of Science (JSPS) KAKENHI (Grant No. 21H05309 and 24K21517) and the Japan Science and Technology Agency (JST), FOREST (Grant No. JPMJFR212N). J.H and C.M. acknowledge funding overseen by the
French National Research Agency (ANR) and France 2030
as part of the Organic Robotics Program (PEPR O2R).


\bibliography{library}

\end{document}